\documentclass[a4paper,11pt]{article}

\usepackage{amsmath}
\usepackage{graphicx}
\usepackage{natbib}
\usepackage{url} 
\usepackage{amsthm}
\usepackage{amssymb}
\usepackage{subfig}
\usepackage{xcolor}%
\usepackage{hyperref}
\usepackage{booktabs}%

\usepackage{enumitem}
\newlist{mydescription}{description}{1}
\setlist[mydescription]{
                        font=\rmfamily\bfseries\upshape,
                        labelindent=\parindent, 
                        leftmargin =2\parindent,
                        rightmargin=\parindent,
                        topsep     =3ex
                       }

\newcommand{\norm}{\mathrm{N}}
\newcommand{\gam}{\mathrm{Gam}}
\newcommand{\expo}{\mathrm{Exp}}
\newcommand{\studt}{\mathrm{t}}

\newcommand{\bdiag}{\mathrm{blockdiag}}
\newcommand{\E}{\mathnormal{E}}
\newcommand{\Var}{\text{Var}}
\newcommand{\Cov}{\text{Cov}}

\newcommand{\diag}{\mathrm{diag}}
\newcommand{\vect}{\mathrm{vec}}
\newcommand{\logit}{\mathrm{logit}}

\newcommand{\trace}{\mathrm{tr}}
\newcommand{\Wishart}{\mathrm{W}}

\def\T{{ \mathrm{\scriptscriptstyle T} }}

\renewcommand{\vec}[1]{\boldsymbol{#1}}
\newcommand{\vecn}[1]{\boldsymbol{#1}}
\newcommand{\matr}[1]{\boldsymbol{#1}}
\newcommand{\matrn}[1]{\boldsymbol{#1}}

\newcommand{\IdLambda}{\matr{\tilde{\Lambda}}}
\newcommand{\idLambda}{\tilde{\lambda}}

\newcommand{\VAR}[1]{\textsc{VAR}$\text{(}{#1}\text{)}$} 

\newcommand{\real}{\mathbb{R}}
\newcommand{\rnth}[1]{\real^{#1}}

\definecolor{five1}{HTML}{F8766D}
\definecolor{five2}{HTML}{A3A500}
\definecolor{five3}{HTML}{00BF7D}
\definecolor{five4}{HTML}{00B0F6}
\definecolor{five5}{HTML}{E76BF3}

\def\dashedrule#1#2#3#4#5#6{{%
	\dimen1=#2 \divide\dimen1 by 2
	\dimen2=#4 \divide\dimen2 by 2
	\def\@ruledash{%
		\rule{\dimen1}{0pt}%
		\rule[0.5ex]{#1}{#6}%
		\rule{\dimen1}{0pt}%
		\rule{\dimen2}{0pt}%
		\rule[0.5ex]{#3}{#6}%
		\rule{\dimen2}{0pt}}%
	\count1=0
	\loop%
	\ifnum\count1<#5%
	\advance\count1 by 1%
	\@ruledash%
\repeat}}

\def\shortsolidline{\dashedrule{1.2em}{0.0em}{0.0em}{0.0em}{1}{0.4pt}}

\def\shortdashedline{\dashedrule{0.3em}{0.1em}{0.0em}{0.0em}{3}{0.4pt}}

\newtheorem{theorem}{Theorem}
\newtheorem{proposition}{Proposition}%

\providecommand{\keywords}[1]
{
  \small
  \textbf{\textit{Keywords---}} #1
}

\title{Structured prior distributions for the covariance matrix in latent factor models}
\author{Sarah E. Heaps and Ian H. Jermyn \\
\small Durham University, Durham, U.K.\\
\small Email: \texttt{sarah.e.heaps@durham.ac.uk}}
\date{}

\begin{document}

\maketitle

\begin{abstract}
Factor models are widely used for dimension reduction in the analysis of multivariate data. This is achieved through decomposition of a $p \times p$ covariance matrix into the sum of two components. Through a latent factor representation, they can be interpreted as a diagonal matrix of idiosyncratic variances and a shared variation matrix, that is, the product of a $p \times k$ factor loadings matrix and its transpose. If $k \ll p$, this defines a parsimonious factorisation of the covariance matrix. Historically, little attention has been paid to incorporating prior information in Bayesian analyses using factor models where, at best, the prior for the factor loadings is order invariant. In this work, a class of structured priors is developed that can encode ideas of dependence structure about the shared variation matrix. The construction allows data-informed shrinkage towards sensible parametric structures while also facilitating inference over the number of factors. Using an unconstrained reparameterisation of stationary vector autoregressions, the methodology is extended to stationary dynamic factor models. For computational inference, parameter-expanded Markov chain Monte Carlo samplers are proposed, including an efficient adaptive Gibbs sampler. Two substantive applications showcase the scope of the methodology and its inferential benefits.
\end{abstract}


\section{\label{sec:introduction}Introduction}

Factor models are widely used as a tool for dimension reduction in explaining the covariance structure of multivariate data. Let $\vec{y}_i = (y_{i1}, \ldots, y_{ip})^\T$ be a $p$-dimensional random vector with mean $\vec{\mu}$ and covariance matrix $\matr{\Omega}$ for observation unit $i=1,\ldots,n$. The classic factor model posits that the covariances between the $p$ components of $\vec{y}_i$ can be explained by their mutual dependence on a smaller number $k$ of unknown \emph{common factors} $\vec{\eta}_i = (\eta_{i1}, \ldots, \eta_{ik})^\T$. Specifically, the model expresses $\vec{y}_i$ as a noisy affine transformation of $\vec{\eta}_i$,
\begin{equation}\label{eq:main}
\vec{y}_i = \vec{\mu} + \matr{\Lambda} \vec{\eta}_i + \vec{\epsilon}_i,
\end{equation}
in which the $p \times k$ matrix $\matr{\Lambda}$ is called the \emph{factor loadings matrix}. The components of the error term $\vec{\epsilon}_i$ are often called the \emph{specific factors} (or ``uniquenesses'') and their variances are often termed the \emph{idiosyncratic variances}. In~\eqref{eq:main}, we take the specific and common factors, $\vec{\epsilon}_i$ and $\vec{\eta}_i$, to be uncorrelated, zero-mean multivariate normal random variables, with
\begin{equation}\label{eq:main2}
\vec{\epsilon}_i  \sim \norm_p(\vecn{0}, \matr{\Sigma}) \quad \text{and} \quad \vec{\eta}_i \sim \norm_k(\vecn{0}, \matr{I}_k).
\end{equation}
In general, the common factors are assumed to explain all the shared variation between the components of $\vec{y}_i$ and so $\matr{\Sigma}$ is constrained to be a diagonal matrix with positive diagonal elements, $\matr{\Sigma}=\diag(\sigma_1^2, \ldots, \sigma_p^2)$. It then follows from~\eqref{eq:main} that the marginal covariance matrix $\matr{\Omega}$ can be expressed as
\begin{equation}\label{eq:omega}
\matr{\Omega} = \matr{\Lambda} \matr{\Lambda}^\T + \matr{\Sigma}
\end{equation}
in which the product $\matr{\Delta} = (\delta_{ij}) = \matr{\Lambda} \matr{\Lambda}^\T$ will henceforth be termed the \emph{shared variation matrix}. \label{pg:rev2_3}If $k \ll p$, the right-hand-side of~\eqref{eq:omega} constitutes a parsimonious factorisation of the full covariance matrix.

Factor models originally found widespread use in psychology and other social sciences \citep[e.g.][]{Gol90}, where a primary motivation was the prospect of domain insight through interpretation of the latent factors. Since their introduction, factor models have been extended in a variety of directions, for example, by allowing sparsity in the matrix of factor loadings for high-dimensional problems \citep[][]{CFHP14} or by modelling the temporal evolution of variances in stochastic volatility models \citep[][]{LC07} or the latent factors in dynamic factor models \citep[][]{SP08}. Accordingly, the use of factor models and their extensions has spread into a diverse array of other fields, such as finance \citep[][]{AW00} and genetics \citep[][]{CCL08}. 

In many of these application areas, the modeller is likely to hold prior beliefs about the nature of the dependence between variables. For example, when the observation vector represents repeated measures data, or observations at a collection of spatial locations, it would be reasonable to expect stronger associations between measurements that are closer together in time or space. Indeed, such ideas underpin the class of functional factor models which are tailored to variables that might be better represented as functions over a continuous domain, rather than vector-valued random quantities. Some functional factor models, and their dynamic extensions, have functional factors \citep[e.g.][]{CVPSK15,TFDB19} while others have functional factor loadings. In the latter case, if we regard the $p$ observations on unit $i$ as measurements on a function over a continuous domain, say $\tau \in \mathbb{R}$, then we replace~\eqref{eq:main} with
\begin{equation*}
y_i(\tau_j) = \mu(\tau_j) + f_i(\tau_j) + \epsilon_i(\tau_j) =  \mu(\tau_j) + \sum_{m=1}^k \lambda_m(\tau_j) \eta_{im} + \epsilon_i(\tau_j)
\end{equation*}
for $j=1,\ldots,p$. The key idea is then to treat the factor loading curves $\lambda_m(\tau)$ as smooth unknown functions of $\tau$. This can be achieved by regarding $f_i(\tau_j) = \sum \lambda_m(\tau_j) \eta_{im}$ as a linear combination of (orthogonal) basis functions, often modelled using splines, which then implies a particular covariance function for $f_i(\tau)$. Although some of the research in this direction uses a Bayesian approach to inference \citep[][]{KMR17,KC22}, the frequentist treatment of the problem is more common, particularly in financial applications to forecasting yield curves \citep[][]{HSH12,JKV14}. An alternative approach in models with functional factor loadings is to assume each $\lambda_m(\tau)$ to be a realisation of a stochastic process, for instance, a Gaussian process with a chosen covariance function; see, for example, the dynamic spatial model of \citet{LSG08}. 

\label{pg:rev2_6a}Beyond analyses of functional data, most of the work in the Bayesian literature on prior specification for factor models has focused on the problem of developing priors that are exchangeable with respect to the order of the variables in the observation vector \citep[e.g.][]{LD16,CLS18} with more recent work focusing additionally on allowing sparsity in the factor loadings matrix \citep[][]{FHL23}. There has been very little work on constructing non-exchangeable prior specifications in a general setting. \label{pg:rev3_1b}A notable exception is the structured increasing shrinkage process prior of \citet{SCD22} under which each factor loading in an infinite factorisation model is assigned a spike-and-slab prior where variable-specific meta-covariates are incorporated in the spike probabilities. \label{pg:rev1_1}Other related work appears in the context of vector error correction models, which are another class of reduced rank models, where \citet{KLS09} develop an informative prior for the matrix of cointegration vectors based on economic theory about the likely cointegration relationships between variables.

In this paper, we treat the factor model as a tool for providing a dimension-reduced parameterisation of a covariance matrix. \label{pg:rev2_4}Based on the insight that the shared variation matrix $\matr{\Delta} = \matr{\Lambda} \matr{\Lambda}^T$ in~\eqref{eq:omega} is more amenable to the specification of prior beliefs than the factor loadings matrix $\matr{\Lambda}$, our main contribution is a framework for incorporating initial beliefs about $\matr{\Delta}$ in the prior for $\matr{\Lambda}$ by exploiting the algebraic relationship between the two quantities. By switching the focus from properties of the prior for $\matr{\Lambda}$ to properties of the prior for $\matr{\Delta}$ we obtain a methodology that is more flexible than alternatives described in the literature. The prior for the shared variation matrix $\matr{\Delta}$ is induced through a structured prior distribution for the factor loadings matrix $\matr{\Lambda}$ which can encode ideas of order-invariance as well as non-exchangeable structure, modelled through dependence on covariance kernels, meta-covariates, pairwise distance metrics and more, within a single principled framework. In particular, the mean of the prior for $\matr{\Delta}$ can be any positive definite matrix of relevance. For example, it could be a phylogenetic distance matrix whose entries encode distances between nodes on a tree, or even the Gram matrix arising from a Gaussian process covariance function. The interpretability of this prior expectation facilitates straightforward incorporation of meaningful domain expertise. Moreover, the flexible class of priors allows data-informed shrinkage towards a region around the mean within a framework that facilitates inference on the number of factors and hence the extent to which dimension-reduction can be achieved. New theoretical results outline the nature of this shrinkage in the rank-reduced space over which $\matr{\Delta}$ has support while shedding light on some of the identifiability issues that can hamper computational inference for factor models.

Based on ideas from the literature on infinite factorisation models and parameter-expansion algorithms, we propose an efficient scheme for computational inference where a single Markov chain Monte Carlo (MCMC) run yields information about both the continuous model parameters and the number of factors. Using an unconstrained reparameterisation of stationary vector autoregressions, the methodology is also extended to a class of stationary dynamic factor models for which the shared variation matrix remains meaningfully defined. To the best of our knowledge, this is the first prior, with associated inferential scheme, for a general class of dynamic factor models that constrains inference to the stationary region without imposing further restrictions. This constitutes an important contribution to the Bayesian time-series literature.

The remainder of this paper is organised as follows. After introducing our parameterisation of Bayesian factor models in Section~\ref{sec:factor_models}, Section~\ref{sec:structured_priors} presents the general class of structured priors and its components. In Section~\ref{sec:dynamic_factor_model}, we extend these ideas to a class of stationary dynamic factor models. Posterior computation is discussed in Section~\ref{sec:posterior_computation}, then Section~\ref{sec:applications} presents a simulation experiment and two substantive applications that illustrate the scope of the methodology and its inferential benefits. 

\section{\label{sec:factor_models}Parameterisation of the Bayesian factor model}

It is well known that the parameters $\matr{\Lambda}$ and $\matr{\Sigma}$ in the marginal covariance matrix $\matr{\Omega}= \matr{\Lambda} \matr{\Lambda}^\T + \matr{\Sigma}$ of a factor model are not identifiable from the likelihood without imposition of constraints. Indeed, even with the scale of the factors fixed at $\Var(\vec{\eta}_i)=\matr{I}_k$ in~\eqref{eq:main2}, there remains a rotational invariance. Consider any $k \times k$ orthogonal matrix $\matr{Q}$. We can pre-multiply the factors in~\eqref{eq:main} by $\matr{Q}$ and post-multiply the factor loadings matrix by $\matr{Q}^\T$ and this gives the transformed factors in~\eqref{eq:main2} the same distribution as the original factors and so the marginal variance in~\eqref{eq:omega} remains unchanged. The factor loadings matrix can therefore only be identified up to an orthogonal transformation. This means there are $pk - k(k-1)/2 + p$ degrees of freedom determining the marginal variance $\matr{\Omega}$. In an unconstrained model for $\vec{y}_i$, the number of degrees of freedom would be $p(p+1)/2$ and so the reduction in the number of parameters is
$p (p+1) / 2 - \{ pk - k (k-1) / 2 + p \}$
which is positive if $k < \varphi(p)$ where
\begin{equation}\label{eq:ledermann_bound}
\varphi(p) = \frac{2p+1-\sqrt{8p+1}}{2}
\end{equation}
is called the Ledermann bound. Notwithstanding rotational invariance, the first identification issue is therefore whether $\matr{\Sigma}$ can be identified uniquely from $\matr{\Omega}$. Fortunately, \citet{BB97} proved that if $k < \varphi(p)$ then $\matr{\Sigma}$, and therefore $\matr{\Delta}=\matr{\Lambda} \matr{\Lambda}^\T$, is almost surely \emph{globally identifiable}. Given identification of $\matr{\Delta}$, solving the rotation problem would then guarantee unique identification of the factor loadings matrix $\matr{\Lambda}$. 

In the Bayesian literature, the most common solution to the rotation problem uses the positive diagonal, lower triangular (PLT) constraint;\label{pg:rev2_5} denoting the constrained matrix by $\IdLambda$ and corresponding factors by $\tilde{\vec{\eta}}_1,\ldots,\tilde{\vec{\eta}}_n$, this demands $\idLambda_{ij}=0$ for $j>i$ and $\idLambda_{ii}>0$ for $i=1,\ldots,k$. Although historically this approach has been widely used in the Bayesian literature \citep[e.g.][]{GZ96,LW04}, we choose not to address the rotational invariance and instead parameterise the model using the unidentified and unconstrained factor loadings matrix $\matr{\Lambda}$. There are two main motivations.

\label{pg:rev1_2}First, recent work provides theoretical and empirical evidence that imposition of the PLT constraint can adversely affect computational inference. Writing $\IdLambda^{(k)}$ to denote the lower triangular matrix comprising the first $k$ rows and columns of $\IdLambda$, problems arise when $\IdLambda^{(k)}$ is near singular, which can occur when variables in $\vec{y}_i$ are highly correlated. When the $j^{\text{th}}$ diagonal element of $\IdLambda^{(k)}$ is close to zero, the signs of $(\idLambda_{j+1,j}, \ldots, \idLambda_{pj})$ are only very weakly identified, producing multiple modes in the posterior corresponding to the different sign flips \citep[e.g.][]{ABP16,MC22}. This can cause poor mixing in MCMC samplers and poor approximation of the posterior for the number of factors \citep[][]{CLS18}. These problems have been addressed successfully by parameter-expansion techniques, which target a posterior over an expanded space with fewer pathological features \citep[e.g.][]{GD09,CLS18}. Parameterising the model and framing the problem of computational inference in terms of the unidentified factor loadings matrix $\matr{\Lambda}$ falls into this class of methods. 

Second, parameterisation in terms of the unconstrained and unidentified factor loadings matrix $\matr{\Lambda}$ delivers an additional benefit in terms of prior specification. \label{pg:rev3_1a}Direct elicitation of a prior for the identified factor loadings in $\IdLambda$ is difficult as they quantify relationships with latent factors whose interpretation is generally unclear \textit{a priori}. In contrast, beliefs about the shared variation $\matr{\Delta}$ in linear Gaussian models can be related directly to beliefs about observable quantities. As a result, $\matr{\Delta}$ is generally a more intuitive parameter for the quantification of covariation. For example, in a spatial problem where the elements of $\vec{y}_i$ correspond to observations at different locations, a modeller might reasonably structure their beliefs through a covariance matrix for which covariance decays as a function of distance. Fortunately, subject to $k < \varphi(p)$, identifiability of $\matr{\Lambda}$ is not necessary for identification of $\matr{\Delta}$. Further, under various standard distributions for an unconstrained random matrix $\matr{\Lambda}$, the first and second order moments of the shared variation matrix $\matr{\Delta}=\IdLambda \IdLambda^\T=\matr{\Lambda}\matr{\Lambda}^\T$ can be calculated in closed form. Through careful specification of the prior for $\matr{\Lambda}$, a modeller can therefore capture their beliefs about the moments in the prior for the shared variation matrix.

\section{\label{sec:structured_priors}Structured prior distributions}

\subsection{\label{subsec:moments}Significance of prior expectation}

The prior expectation of the shared variation matrix $\matr{\Delta}=\matr{\Lambda} \matr{\Lambda}^\T$ is defined by $\E(\matr{\Delta}) = \{ \E(\delta_{ij}) \} = \E(\matr{\Lambda} \matr{\Lambda}^\T)$. As we detail in Section~\ref{subsec:prior_for_xi}, $\E(\matr{\Delta})$ will be chosen to reflect beliefs about the covariation amongst the $p$ elements of $\vec{y}_i$. In general, it will be a rank $p$ matrix.

For $k < p$, denote by $\mathcal{S}_{p,k}^+$ the set of rank $k$, $p \times p$ symmetric, \label{pg:rev1_3}positive semi-definite matrices and write $\mathcal{S}_{p}^+$ for the space of $p \times p$ symmetric, positive definite matrices. The factor loadings matrix $\matr{\Lambda}$ is rank $k < \varphi(p) < p$ and so the prior distribution for the shared variation matrix $\matr{\Delta}=\matr{\Lambda}\matr{\Lambda}^\T$ only offers non-zero support to $\mathcal{S}_{p,k}^+$. Because $\mathcal{S}_{p,k}^+$ is not a convex set, it need not contain the prior expectation of $\matr{\Delta}$. Indeed, as stated previously, $\E(\matr{\Delta})$ will generally be rank $p$. Therefore, although $\E(\matr{\Delta})$ represents the centre of prior mass, in general there will be zero density at this point. The significance of $\E(\matr{\Delta})$ is thus not completely clear. We will elucidate its meaning via an alternative, constrained expectation, as follows.

The Frobenius norm of a matrix is simply the Euclidean norm of its vectorised form. Define the (squared) \emph{Frobenius distance} between two matrices $\matr{A}$ and $\matr{B}$ as the squared Frobenius norm of their difference:
\begin{equation*}
d(\matr{A}, \matr{B})^2 = \| \matr{A} - \matr{B} \|_F^2 = \trace\left\{ (\matr{A} - \matr{B})^\T (\matr{A} - \matr{B}) \right\}.
\end{equation*}
By analogy with the classic Euclidean expectation, we now define the \emph{constrained expectation} of $\matr{\Delta}$ as $\E_F(\matr{\Delta}) = \matr{L}_0 \matr{L}_0^\T \in \mathcal{S}_{p,k}^+$ where
\begin{equation*}
\matr{L}_0 = \min_{\matr{L} \in \mathbb{R}^{p \times k}} \E_{\Lambda}  \left\{ d(\matr{\Lambda} \matr{\Lambda}^\T, \matr{L} \matr{L}^\T)^2 \right\}.
\end{equation*}
Theorem~\ref{thm:mean_is_min} below asserts that if the $k^{\text{th}}$ and $(k+1)^{\text{th}}$ eigenvalues of $\E(\matr{\Delta})$ are distinct, the constrained expectation of $\matr{\Delta}$ is the point in $\mathcal{S}_{p,k}^+$ that minimises the Frobenius distance to $\E(\matr{\Delta})$. However, before proving Theorem~\ref{thm:mean_is_min}, we need Proposition~\ref{prop:min_distance}.

\begin{proposition}\label{prop:min_distance}
Let the spectral decomposition of a matrix $\matr{D} \in \mathcal{S}_p^+$ be $\matr{U} \matr{M} \matr{U}^\T$, where $\matr{M} = \diag(m_1, \ldots, m_p)$ is a diagonal matrix of ordered eigenvalues, $m_1 \ge m_2 \ge \cdots \ge m_p > 0$, and $\matr{U}$ is an orthogonal matrix whose columns comprise the corresponding eigenvectors. Assume that $m_k \ne m_{k+1}$. Then, for $\matr{\Lambda} \in \mathbb{R}^{p \times k}$, the matrix product $\matr{\Lambda} \matr{\Lambda}^\T$ which minimises the Frobenius distance to $\matr{D}$ is $\matr{\Lambda} \matr{\Lambda}^\T = \matr{D}^{1/2} \matr{U}^{(k)} \matr{U}^{(k)\, \T} \matr{D}^{1/2}$ where $\matr{U}^{(k)}$ is the sub-matrix comprising the first $k$ columns of $\matr{U}$. Moreover, the minimum squared Frobenius distance is equal to the sum of the squares of the last $p-k$ eigenvalues of $\matr{D}$.
\end{proposition}

The proof of Proposition~\ref{prop:min_distance} is provided in the Supplementary Materials.

\begin{theorem}\label{thm:mean_is_min}
If $\matr{U} \matr{M} \matr{U}^\T$ denotes the spectral decomposition of $\E(\matr{\Delta}) \in \mathcal{S}_p^+$ and $m_k \ne m_{k+1}$, then the constrained expectation of $\matr{\Delta}$, $\E_F(\matr{\Delta}) \in \mathcal{S}_{p,k}^+$, is equal to the matrix product which minimises the Frobenius distance to $\E(\matr{\Delta})$. That is, $\E_F(\matr{\Delta}) = \E(\matr{\Delta})^{1/2} \matr{U}^{(k)} \matr{U}^{(k)\, \T} \E(\matr{\Delta})^{1/2}$.
\end{theorem}

The proof of Theorem~\ref{thm:mean_is_min} is given in the Supplementary Materials. Its significance lies in the suggestion that the prior for $\matr{\Delta}$ encourages shrinkage towards the closest matrix in $\mathcal{S}_{p,k}^+$ to the rank $p$ prior expectation $\E(\matr{\Delta})$, hence for a given structure for $\E(\matr{\Delta})$, approximating it as closely as possible in rank-reduced form. The Supplementary Materials also consider the case $m_k = m_{k+1}$ and the implications for computational inference.

\subsection{\label{subsec:matrix_normal_prior}A matrix normal prior}

A random matrix $\matr{\Lambda}$ has a matrix normal distribution with location matrix $\matr{M} \in \mathbb{R}^{p \times k}$, among-row scale matrix $\matr{\Phi} \in \mathcal{S}_p^+$ and among-column scale matrix $\matr{\Psi} \in \mathcal{S}_k^+$, written $\matr{\Lambda} \sim \norm_{p,k}(\matr{M}, \matr{\Phi}, \matr{\Psi})$, if 
$\vect(\matr{\Lambda})$ is multivariate normal and such that $\vect(\matr{\Lambda}) \sim \norm_{pk} \Bigl( \vect(\matr{M}), \matr{\Psi} \otimes \matr{\Phi} \Bigr)$. Since $(\alpha \matr{\Psi}) \otimes (\alpha^{-1} \matr{\Phi}) = \matr{\Psi} \otimes \matr{\Phi}$ for any scalar $\alpha > 0$, the overall scale of either $\matr{\Phi}$ or $\matr{\Psi}$ can be fixed without loss of generality.  

Suppose that we take $\matr{\Lambda} \sim \norm_{p,k}(\matrn{0}, \matr{\Phi}, \matr{\Psi})$ as a prior for the unconstrained factor loadings matrix and fix the scale of the among-row scale matrix by taking $\trace(\matr{\Phi})=p$ or $\trace(\matr{\Phi}^{-1})=p$. As remarked in Section~\ref{sec:factor_models}, the shared variation matrix $\matr{\Delta} = (\delta_{ij}) = \matr{\Lambda} \matr{\Lambda}^\T$ is a quantity which is amenable to the specification of prior beliefs, and so our goal is to derive its moments. Using standard theory for the matrix normal distribution \citep[e.g.][Chapter 2]{GN00}, it can be shown that the expectation of $\matr{\Delta}$ is given by
\begin{equation}\label{eq:matrix_norm_mean}
\E(\matr{\Delta}) = \trace(\matr{\Psi}) \matr{\Phi},
\end{equation}
while the covariance between $\delta_{ij}$ and $\delta_{k\ell}$ is $\Cov(\delta_{ij}, \delta_{k\ell}) = \trace(\matr{\Psi}^2) (\phi_{ik} \phi_{j\ell} + \phi_{i\ell} \phi_{jk})$. In the special case where $i=k$ and $j=\ell$, the variance of $\delta_{ij}$ is then 
\begin{equation}\label{eq:matrix_norm_var}
\Var(\delta_{ij}) = \trace(\matr{\Psi}^2) (\phi_{ii} \phi_{jj} + \phi_{ij}^2).
\end{equation}
The derivations of the moments above are provided in the Supplementary Materials.

The result in~\eqref{eq:matrix_norm_mean} is significant because of the interpretation it bestows on the among-row scale matrix $\matr{\Phi}$ as a standardised version of our prior expectation for the shared variation matrix $\matr{\Delta}$. Parametric forms are convenient models for covariance matrices because they provide a parsimonious representation of dependence and a way to model the relationships with as-yet unobserved variables. We can therefore model the among-row scale matrix $\matr{\Phi}$ using a parametric form. This encourages shrinkage of the shared variation matrix $\matr{\Delta}$ towards an area around that parametric form but without enforcing the structure as a model constraint. 

Suppose that $\matr{\Phi}=\matr{R}(\vec{\vartheta})$ in which $\vec{\vartheta}$ is a low-dimensional vector of hyperparameters on which the parametric form $\matr{R}(\cdot)$ depends. We complete our prior for the factor loadings matrix by taking $\matr{\Psi}=\diag(\psi_1, \ldots, \psi_k)$ and assigning a prior to $\vec{\vartheta}$ and the $\psi_i$. Parametric forms $\matr{R}(\cdot)$ and the prior for the $\psi_i$ are discussed in Sections~\ref{subsec:prior_for_xi} and~\ref{subsec:unknown_k}, respectively.

\subsection{\label{subsec:matrix_t_prior}A matrix-$t$ prior}

The expressions for the means and variances in~\eqref{eq:matrix_norm_mean} and~\eqref{eq:matrix_norm_var} under the matrix normal prior reveal a lack of flexibility; recalling that the scale of $\matr{\Phi}$ is fixed, once the mean $\E(\matr{\Delta})$ has been chosen, it is clearly not possible to scale up or down all the variances by scaling $\matr{\Psi}$. 
As such, we cannot assess independently the overall scale of the mean and uncertainty around that mean. A matrix-$t$ prior for $\matr{\Lambda}$ remedies this problem through the introduction of a degree of freedom parameter.

Let $\matr{S} \sim \Wishart_p(\varsigma+p-1, \matr{\Phi}^{-1})$ and $\matr{X} \sim \norm_{p,k}(\matrn{0}, \matr{I}_p, \matr{\Psi})$ be independent random matrices, where $\Wishart_n(q, \matr{U})$ denotes the $n$-dimensional Wishart distribution with $q>0$ degrees of freedom and scale matrix $\matr{U} \in \mathcal{S}_n^+$. If we define
\begin{equation}\label{eq:representation_matrixt}
\matr{\Lambda} = (\matr{S}^{-1/2})^\T \matr{X} + \matr{M}
\end{equation}
where $\matr{S}^{1/2} (\matr{S}^{1/2})^\T=\matr{S}$ and $\matr{M} \in \mathbb{R}^{p \times k}$, then the random matrix $\matr{\Lambda}$ has a matrix-$t$ distribution, written $\matr{\Lambda} \sim \studt_{p,k}(\varsigma, \matr{M}, \matr{\Phi}, \matr{\Psi})$. Again, one can fix the scale of either the among-row scale matrix $\matr{\Phi} \in \mathcal{S}_p^+$ or the among-column scale matrix $\matr{\Psi} \in \mathcal{S}_k^+$ without loss of generality.

Suppose that we adopt $\matr{\Lambda} \sim \studt_{p,k}(\varsigma, \matrn{0}, \matr{\breve{\Phi}}, \matr{\Psi})$ as a prior for the factor loadings matrix. Expressions for the means, variances and covariances of $\matr{\Delta}=\matr{\Lambda} \matr{\Lambda}^\T$ are derived in the Supplementary Materials. In particular, we show that $\E(\matr{\Delta})=\trace(\matr{\Psi}) \matr{\Phi}$ for $\varsigma>2$ where $\matr{\Phi} = \matr{\breve{\Phi}} / (\varsigma-2)$, which is identical to the expression derived under the matrix normal prior. We also show that for a fixed value of $\E(\matr{\Delta})$, the variance,
\begin{equation*}
\Var(\delta_{ij}) = \frac{\left\{ \trace(\matr{\Psi})^2 + (\varsigma-2) \trace(\matr{\Psi}^2) \right\} \left\{ \varsigma \phi_{ij}^2 + (\varsigma-2) \phi_{ii} \phi_{jj} \right\}}{(\varsigma-1)(\varsigma-4)} \quad \text{for $\varsigma>4$,}
\end{equation*}
can be increased or decreased by decreasing or increasing $\varsigma$, respectively. Therefore, by treating $\varsigma>4$ as an unknown and assigning it a prior, we can allow the data to influence the degree of shrinkage of the shared variation matrix towards the closest rank $k$ matrix to its mean. Writing $\check{\varsigma} = 1 / (\varsigma-4)  \in \mathbb{R}^+$, a matrix normal distribution for $\matr{\Lambda}$ is recovered as $\check{\varsigma} \to 0$. We can therefore allow controlled relaxation of the fixed shrinkage of the matrix normal prior by specifying a prior for $\check{\varsigma}$ with its mode at zero. \label{pg:rev1_4}To this end, we recommend an exponential distribution $\check{\varsigma} \sim \expo(a_0)$. In the special case when $\matr{\Phi}=\matr{I}_p$ and $\matr{\Psi}=\psi \matr{I}_k$, we have $\Var(\delta_{ij}) = s_k(\check{\varsigma}) \E(\delta_{ij})$ where $s_k(\check{\varsigma}) = (1+2\check{\varsigma}) \left\{ 1 + (2+k)\check{\varsigma} \right\} / (1+3\check{\varsigma})$ is an increasing function in $\check{\varsigma}$ which takes its minimum value of 1 when $\check{\varsigma}=0$. By trial and improvement, we can therefore use the quantiles of the $\expo(a_0)$ distribution to choose a value for $a_0$ which is consistent with our prior beliefs about a chosen quantile in the distribution for the scale factor $s_k(\check{\varsigma})$ for various $k$. This is illustrated in the application in Section~\ref{subsec:finnish_birds}.

\subsection{\label{subsec:prior_for_xi}Model and prior for $\matr{\Phi}$}

\subsubsection{General form}
Under either a matrix normal or matrix-$t$ prior for $\matr{\Lambda}$, $\E(\matr{\Delta} | \matr{\Phi}, \matr{\Psi})=\trace(\matr{\Psi}) \matr{\Phi}$. As explained in Section~\ref{subsec:matrix_normal_prior}, since we choose to fix $\trace(\matr{\Phi})=p$ or $\trace(\matr{\Phi}^{-1})=p$, we can therefore interpret $\matr{\Phi}$ as a standardised version of the prior expectation of $\matr{\Delta}$. For most of the parametric forms described in this section, it will be possible to factorise $\matr{\Phi}$ or $\matr{\Xi}=\matr{\Phi}^{-1}$ as $\matr{\Phi}=\matr{R}(\vec{\vartheta})$ or $\matr{\Xi}=\matr{R}(\vec{\vartheta})$, respectively, where $\matr{R}(\cdot)$ yields a positive definite matrix with 1s on the diagonal. The vector $\vec{\vartheta}$ typically contains a small number of unknown hyperparameters to which we assign a prior. 

\subsubsection{Specific examples}
The simplest structure for $\matr{\Phi}$ would take $\matr{\Phi} = \matr{I}_p$ giving $\E(\matr{\Delta} | \matr{\Phi}, \matr{\Psi})=\trace(\matr{\Psi}) \matr{I}_p$. In the matrix normal case when, additionally, $\matr{\Psi}=\psi \matr{I}_k$, we obtain the order-invariant prior presented in \citet{LD16} for the identified factor loadings matrix in which $\idLambda_{ij} \sim \norm(0, \psi)$ for $i \ne j$ and $\idLambda_{ii}^2 \sim \gam\{(k-i+1)/2, 1 / (2 \psi)\}$ for $i=1,\ldots,k$. 
Although taking $\matr{\Phi}=\matr{I}_p$ will give an order-invariant prior for $\matr{\Delta}$, a more general distribution that is exchangeable with respect to the order of the variables in $\vec{y}_i$ arises by taking $\matr{\Phi}$ to be a two-parameter exchangeable matrix. Since we constrain $\trace(\matr{\Phi})=p$ this yields $\matr{\Phi}=(1 - \vartheta) \matr{I}_p + \vartheta \matr{J}_p$, where $-1/(p-1) < \vartheta < 1$, $\matr{J}_p = \vecn{1}_p \vecn{1}_p^\T$ and $\vecn{1}_p$ is a $p$-vector of 1s. This is the most general form for a $p \times p$ symmetric, positive definite matrix with $\trace(\matr{\Phi})=p$ which is invariant under a common permutation of the rows and columns. If a modeller has nothing in their prior beliefs to distinguish among the $p$ elements of $\vec{y}_i$, the advantage of this specification over the order-invariant prior of \citet{LD16} is that it allows shrinkage of the $p(p-1)/2$ off-diagonal elements of the shared variation matrix $\matr{\Delta}$ towards non-zero values. This can be particularly helpful when the number of observations is small relative to the dimension of $\vec{y}_i$.

In confirmatory factor analysis, structural zeros are often introduced in the factor loadings matrix based on prior beliefs about the relationships between the $k$ factors and the variables in $\vec{y}_i$. For example, in the prototypical two-factor model, the first $p_1$ variables load only on factor 1 whilst the remaining $p-p_1$ variables load only on factor 2. This would correspond to a block diagonal shared variation matrix $\matr{\Delta}$. Rather than imposing these beliefs about independent group structure as a model constraint, one could instead encourage shrinkage towards the corresponding structure through the prior. With appropriate ordering of the variables, this could be achieved by taking $\matr{\Phi} = \bdiag(\matr{\Phi}_1, \ldots, \matr{\Phi}_k)$ in which each $\matr{\Phi}_i$ block is a two-parameter exchangeable matrix, as described above.

Beyond these simple examples, there are many fields of statistics in which parametric forms are adopted for correlation matrices or their inverses. Typically, they are intended to capture the idea that observations from variables that are closer together, often in time or space, tend to be more strongly (positively) correlated.  For instance, the precision matrix for a vector of longitudinal measurements might be assumed to be that of a stationary autoregressive process of order one. We could centre the prior for $\matr{\Delta}$ on a matrix of this form by taking $\matr{\Xi}=(\xi_{ij})$ to be a tridiagonal matrix with $\xi_{i,i+1}=\xi_{i+1,i}=-\vartheta$ for $i=1,\ldots,p-1$, $\xi_{ii}=1+\vartheta^2$ for $i=2,\ldots,p-1$ and $\xi_{11}=\xi_{pp}=1$ where $\vartheta \in (-1, 1)$. 

In spatial statistics, it is very common to use a spatial correlation function, such as one from the Mat{\'e}rn class. A conventional spatial correlation function will depend only on the distance, and possibly direction, between sites. However, \citet{SGO11} introduced the idea of recording distances in a space of dimension greater than two, by introducing covariates into the correlation function. A special case, termed the projection model, measures the distance between sites $j$ and $k$ in this augmented space using the Mahalanobis metric with respect to an (unknown) $c \times c$ symmetric, positive definite matrix $\matr{\Theta}$, that is,
\begin{equation}\label{eq:distance}
d_{C,jk} = \surd \{ (\vec{x}_{j}-\vec{x}_{k})^{\T} \matr{\Theta} (\vec{x}_{j}-\vec{x}_{k}) \}.
\end{equation}
Here $\vec{x}_j$ is a vector of coordinates for site $j$, comprising its two spatial coordinates and the values of $c-2$ meta-covariates. This is then combined with an exponential spatial correlation function to give $\phi_{jk} = \exp(-d_{C,jk})$. Omitting the spatial coordinates, these ideas can be extended to non-spatial settings if there are meta-covariates associated with the $p$ variables that are thought to influence the relationships between deviations from the mean $(y_{ij} - \mu_j)$ and $(y_{ik} - \mu_k)$ in any vector measurement. 
Indeed, any Gaussian process correlation matrix could be used for $\matr{\Phi}$ which has huge scope for the incorporation of meaningful domain expertise.

More generally, if the distances between variables can be measured by some metric, say $d_{jk}$, this could be used to structure the prior through the specification $\phi_{jk} = \exp(-d_{jk} / \vartheta)$. An ecological example is provided in Section~\ref{subsec:finnish_birds} which uses phylogenetic distances between species in a joint model for species occurrence.

\subsection{\label{subsec:unknown_k}Model and prior for $\matr{\Psi}$}

Except in confirmatory factor analysis, an appropriate number of factors is not usually known \textit{a priori}. Various methods have been proposed in the literature to address this source of uncertainty. This includes techniques to approximate the marginal likelihood of models with different numbers of factors, including path sampling \citep[][]{DG13} and Chib's method \citep[][]{CNS06}. However, these methods are very computationally expensive, requiring separate Monte Carlo simulations for models with $k=1,\ldots,H$ factors, where $H$ is the maximum number of factors that a modeller is prepared to entertain. \citet{LW04} propose a reversible jump MCMC algorithm but the proposals for transdimensional moves require pilot MCMC runs for models with $k=1,\ldots,H$ factors which rapidly becomes untenable for high-dimensional problems. \label{pg:rev2_6b}Other methods that attempt to allow simultaneous inference of the number of factors and the model parameters rely on identification of individual zeros in the factor loadings matrix \citep[][]{FL10,CFHP14,FHL23}. The approach taken here in similar in spirit, relying on the removal of columns whose contribution to $\matr{\Omega}$ is negligible.

Due to the challenges in approximating the posterior for $k$, a recent approach that has become popular relies conceptually on an overfitted factor model \citep[][]{BD11,LDD20,SCD22}. In theory, infinitely many factors are permitted but cumulative shrinkage priors are used to increasingly shrink columns of the factor loadings matrix towards zero as the column index increases. This allows the infinite factor loadings matrix to be approximated with a finite matrix by omitting columns that have been shrunk close enough to zero. The posterior for the number of non-discarded columns $k^{\ast}$ is then used as a proxy for the posterior for $k$. As such, a single MCMC sample yields information about both the model parameters and the number of latent factors. 

A popular increasing shrinkage prior is the multiplicative gamma process (MGP) \citep[][]{BD11}. Conditional on a set of hyperparameters, the factor loadings within each column are assumed to be independent normal random variables with zero mean and a column-specific precision parameter. The precisions are a cumulative product of gamma random variables and their prior expectation increases with the column index. A factor is deemed inactive \textit{a posteriori} if the contribution of its loadings to $\matr{\Omega}$ is negligible under a chosen truncation criterion. In order to apply these ideas to our structured prior for the shared variation matrix, we take the among-column scale matrix $\matr{\Psi}$ to be diagonal and assign the reciprocals of its elements a MGP prior
\begin{equation}\label{eq:mgp}
\psi_h^{-1} = \prod_{\ell=1}^h \varrho_{\ell}, \quad \varrho_{1} \sim \gam(a_1, 1), \quad \varrho_{\ell} \sim \gam(a_2, 1), \quad \ell \ge 2
\end{equation}
in which the $\varrho_{\ell}$ are independent. To ensure identifiability of the shared variation matrix $\matr{\Delta}=\matr{\Lambda} \matr{\Lambda}^\T$ and the idiosyncratic variances $\matr{\Sigma}$ in the likelihood, we choose to truncate the prior at $H \le \lceil \varphi(p) \rceil - 1$ where $\varphi(p)$ was defined in~\eqref{eq:ledermann_bound}. Guidelines on the choice of $a_1$ and $a_2$ to ensure the $1/\psi_h$ are stochastically increasing in a reasonable neighbourhood of zero are provided in \citet{Dur16}.

\section{\label{sec:dynamic_factor_model}Stationary dynamic factor models}

\subsection{Bayesian dynamic factor models}
In the classic vector autoregressive (VAR) model, the number of parameters grows quadratically with the dimension of the observation vector. In contrast, the number of parameters in a dynamic factor model only grows quadratically with the dimension of the latent factors. This makes them a valuable tool for dimension-reduction in multivariate time series analysis. A dynamic factor model has the same observation equation as a static model,
\begin{equation}\label{eq:dynamic_factor_model_obs}
\vec{y}_t = \vec{\mu} + \matr{\Lambda} \vec{\eta}_t + \vec{\epsilon}_t, \qquad \vec{\epsilon}_t \sim \norm_p(\vecn{0}, \matr{\Sigma}).
\end{equation}
However, the factors in a dynamic model evolve over time according to an order $m$ vector autoregression (\VAR{m}). This yields the state equation as
\begin{equation}\label{eq:dynamic_factor_model_fac}
\vec{\eta}_t = \matr{\Gamma}_1 \vec{\eta}_{t-1} + \ldots + \matr{\Gamma}_m \vec{\eta}_{t-m} + \vec{\zeta}_t, \qquad \vec{\zeta}_t \sim \norm_k(\vecn{0}, \matr{\Pi}),
\end{equation}
for $t=1,2,\ldots$ in which the observation and factor innovations $\vec{\epsilon}_t$ and $\vec{\zeta}_t$ are serially and mutually uncorrelated. As in the static case, it is generally assumed that the common factors explain all of the shared variation and so $\matr{\Sigma}$ is taken to be diagonal. Various extensions of this model have been presented, for example, \citet{PP06} incorporate a moving average component in state equation, \citet{ABP16} introduce lagged factors in~\eqref{eq:dynamic_factor_model_obs} and \citet{KS19} allow the components of the idiosyncratic innovations to evolve independently as univariate stationary autoregressions.

\subsection{A stationary dynamic factor model}
The methodology described in this paper relies on the use of a structured prior distribution for the shared variation component of the marginal variance. The marginal variance of a dynamic factor model is only meaningfully defined when the factors evolve according to a stationary process; hereafter we focus only on situations where this is a reasonable assumption. The manner by which stationarity is enforced is discussed in Section~\ref{subsec:enforcing_stationarity}. 

In order to identify the scale of the loadings matrix in a dynamic factor model, the scale of the factors is normally constrained by fixing $\matr{\Pi}=\Var(\vec{\zeta}_t)=\matr{I}_k$. However, to retain the interpretation of $\matr{\Delta}=\matr{\Lambda}\matr{\Lambda}^\T$ as a shared variation matrix, we instead constrain the stationary covariance matrix so that $\matr{G}_0=\Var(\vec{\eta}_t)=\matr{I}_k$ and hence the marginal variance of $\vec{y}_t$ remains as $\matr{\Omega}=\matr{\Lambda} \matr{\Lambda}^\T + \matr{\Sigma}$. We can therefore assign a prior to the unconstrained factor loadings matrix $\matr{\Lambda}$ using precisely the ideas discussed in Section~\ref{sec:structured_priors} for the static model. 

To complete specification of the stationary dynamic factor model, the state equation~\eqref{eq:dynamic_factor_model_fac} must be initialised at its stationary distribution. To this end, we augment the parameter space with $m$ auxiliary factors at times $t=1-m, \ldots, 0$. Generalising the definition of the stationary variance $\matr{G}_0$ above, we define $\matr{G}_i = \Cov(\vec{\eta}_t, \vec{\eta}_{t+i})$ as the $i^{\text{th}}$ autocovariance of $\vec{\eta}_t$. The initial distribution can then be expressed as $(\vec{\eta}_{1-m}^\T, \ldots, \vec{\eta}_0^\T)^\T \sim \norm_{km}(\vecn{0}, \matr{G})$ where $\matr{G}$ is a positive definite block Toeplitz matrix with $\matr{G}_{j-i}$ as the block in rows $\{k(i-1)+1\}$ to $ki$ and columns $\{k(j-1)+1\}$ to $kj$ ($i,j=1,\ldots,m$) and $\matr{G}_{-\ell}=\matr{G}_\ell^\T$ ($\ell=1,\ldots,m-1$).


\subsection{\label{subsec:enforcing_stationarity}Enforcing the stationarity constraint}
Denote by $B$ the backshift operator, $B \vec{\eta}_t = \vec{\eta}_{t-1}$. The VAR model governing evolution of the factors can then be written as $\vec{\zeta}_t = (\matr{I}_k - \matr{\Gamma}_1 B - \cdots - \matr{\Gamma}_m B^m) \vec{\eta}_t = \matr{\Gamma}(B) \vec{\eta}_t$ where $\matr{\Gamma}_i \in \mathbb{R}^{k \times k}$ for $i=1,\ldots,m$ and $\matr{\Gamma}(u) = \matr{I}_k - \matr{\Gamma}_1 u - \cdots - \matr{\Gamma}_m u^m$, $u \in \mathbb{C}$, is termed the characteristic polynomial. The process is stable if and only if all the roots of $\det \{ \matr{\Gamma}(u) \} = 0$ lie outside the unit circle. Because all stable processes are stationary and unstable stationary processes are not generally of interest, this subset of the Cartesian product space in which the $\matr{\Gamma}_i$ lie is often referred to as the stationary region, $\mathcal{C}_{m, k}$. Even for the simplest vector case of an order-1 bivariate autoregression, it has a highly complex geometry.

Various authors have considered Bayesian inference for stationary dynamic factor models \citep[e.g.][]{SP08,LSG08}. However, due to the difficulties of designing efficient MCMC samplers with state space constrained to $\mathcal{C}_{m, k}$, the autoregressive coefficient matrices are often assumed to be diagonal, which simplifies the stationarity condition to that of a univariate autoregression. Fortunately, recent work by \citet{Hea22} presents a prior for the autoregressive coefficient matrices that is constrained to $\mathcal{C}_{m, k}$ and facilitates routine computational inference. This is based on an unconstrained reparameterisation of the model through two bijective transformations. First, the model parameters $\{ (\matr{\Gamma}_1, \ldots, \matr{\Gamma}_m), \matr{\Pi} \}$ undergo a recursive mapping which yields a new parameter set $\{ (\matr{P}_1, \ldots, \matr{P}_m), \matr{\Pi} \}$. Here $\matr{P}_{i+1}$ is the $(i+1)^{\text{th}}$ partial autocorrelation matrix, that is, a standardised version of the conditional cross-covariance between $\vec{\eta}_{t+1}$ and $\vec{\eta}_{t-i}$ given the $i$ intervening variables $(\vec{\eta}_{t}, \ldots, \vec{\eta}_{t-i+1})$. Each partial autocorrelation matrix lies in the space of $k \times k$ square matrices with singular values less than one. By mapping the singular values of $\matr{P}_i$ from $[0,1)$ to the positive real line, a second mapping then constructs an unconstrained $k \times k$ square matrix $\matr{A}_i$ through $\matr{A}_i=(\matr{I}_k - \matr{P}_i \matr{P}_i^\T)^{-1/2} \matr{P}_i$, $i=1,\ldots,m$, in which $\matr{X}^{-1/2}$ denotes the inverse of the symmetric matrix square root of $\matr{X}$. Specification of a prior for the $\matr{A}_i$, discussed in Section~\ref{subsec:dyn_param_expansion}, and computational inference is now routine owing to the Euclidean geometry of the parameter space.

The inverse mapping from the intermediate reparameterisation $\{ (\matr{P}_1, \ldots, \matr{P}_m), \matr{\Pi} \}$ to the original parameters $\{ (\matr{\Gamma}_1, \ldots, \matr{\Gamma}_m), \matr{\Pi} \}$ involves two recursions. The first outputs the stationary covariance matrix $\matr{G}_0$ from $\{ (\matr{P}_1, \ldots, \matr{P}_m), \matr{\Pi} \}$ and the second outputs $(\matr{\Gamma}_1, \ldots, \matr{\Gamma}_m)$ (and recovers $\matr{\Pi}$) from $(\matr{P}_1, \ldots, \matr{P}_m)$ and $\matr{G}_0$. It is therefore trivial to modify the reparameterisation to accommodate the constraint that $\matr{G}_0=\matr{I}_k$; one simply omits the first recursion, and calculates $\{ (\matr{\Gamma}_1, \ldots, \matr{\Gamma}_m), \matr{\Pi} \}$ from $(\matr{P}_1, \ldots, \matr{P}_m)$ and $\matr{G}_0=\matr{I}_k$ in the second recursion. We note that the autocovariance matrices $\matr{G}_1, \ldots, \matr{G}_{m-1}$ needed to characterise the initial distribution of $\vec{\eta}_{(1-m):0}$ are also by-products of this second recursion.

When $\vec{\eta}_t$ follows a \VAR{1} process, there is a closed form expression for the original model parameters in terms of $\matr{P}_1$ or, equivalently, $\matr{A}_1$. Dropping the 1-subscript for brevity, we can write $\matr{\Gamma}=\matr{P}=(\matr{I}_k+\matr{A}\matr{A}^\T)^{-1/2} \matr{A}$ and $\matr{\Pi}=\matr{I}_k-\matr{P} \matr{P}^\T=(\matr{I}_k+\matr{A} \matr{A}^\T)^{-1}$.

\subsection{\label{subsec:dyn_param_expansion}Prior for the transformed partial autocorrelation matrices}

The stationary dynamic factor model described in the previous sections is invariant under rotations of the factors when compensatory transformations are applied to the factor loadings and parameters of the state equation. In particular, marginalising over the common factors, it can readily be shown that the likelihood evaluated at $(\matr{\mu},\matr{\Lambda},\matr{\Sigma},\matr{A}_1,\ldots,\matr{A}_m)$ is the same as the likelihood evaluated at $(\matr{\mu},\matr{\Lambda} \matr{Q}^\T,\matr{\Sigma},\matr{Q} \matr{A}_1 \matr{Q}^\T,\ldots,\matr{Q} \matr{A}_m \matr{Q}^\T)$ for any orthogonal matrix $\matr{Q} \in \mathcal{O}(k)$. In the absence of any meaningful information to distinguish $\matr{A}_i$ from $\matr{Q} \matr{A}_i \matr{Q}^\T$ \textit{a priori}, we assign a prior which is rotatable, that is $\matr{A}_i$ and $\matr{Q} \matr{A}_i \matr{Q}^\T$ have the same distribution for any orthogonal matrix $\matr{Q}$. To this end, we take the $\matr{A}_i$ to be independent with $\matr{A}_i \sim \norm_{k,k}(\matrn{0}, \matr{I}_k, \matr{I}_k)$, $i=1,\ldots,m$.

\section{\label{sec:posterior_computation}Posterior computation}

\subsection{Samplers with fixed truncation level}

Consider first the static factor model. For computational inference, we use MCMC to sample from the posterior associated with our unconstrained parameterisation of the model. When the truncation level in the factor loadings matrix is fixed at $H$, we propose a straightforward Gibbs sampler. The full conditional distributions for most unknowns have standard forms and can be sampled directly. However, depending on the structure chosen for the among-row scale matrix $\matr{\Phi}=\matr{R}(\vec{\vartheta})$, a semi-conjugate prior for the correlation parameter(s) in $\vec{\vartheta}$ is not generally available and so we update $\vec{\vartheta}$ in a Metropolis-Hastings step. Computational inference is slightly complicated if a matrix-$t$, rather than matrix normal, prior is adopted for $\matr{\Lambda}$ because it is not conjugate to the likelihood. However, sampling becomes straightforward if we make use of the representation of a matrix-$t$ random variable in~\eqref{eq:representation_matrixt} and augment the state space of the sampler with the matrix $\matr{S}$. The full conditional distribution for $\matr{\Lambda}$ is then matrix-normal while the full conditional distribution for $\matr{S}$ is Wishart. The only other additional unknown is the degree of freedom parameter $\varsigma$. If $\varsigma$ is assigned the prior from Section~\ref{subsec:matrix_t_prior}, its full conditional distribution is non-standard. Rather than updating the transformed degree of freedom parameter $\check{\varsigma}$ through its own Metropolis-Hastings step, mixing can be greatly improved by performing a joint update of $\matr{S}$ and $\check{\varsigma}$. The proposal density takes the form $q(\matr{S}^\ast, \check{\varsigma}^\ast | \matr{S}, \check{\varsigma},\ldots)=q_1(\check{\varsigma}^\ast | \check{\varsigma})q_2(\matr{S}^\ast | \check{\varsigma}^\ast,\ldots)$ in which $q_1(\cdot)$ is a random walk proposal on the log-scale and $q_2(\cdot)$ is the full conditional density for $\matr{S}$. 

Extension of the Gibbs sampler to handle the dynamic factor model is straightforward. \label{pg:rev1_5}The latent factors $\vec{\eta}_{1-m}, \ldots, \vec{\eta}_n$ can be sampled efficiently from their full conditional distribution in a single block using a variety of sampling approaches \citep[][]{WY02,WY04}. \label{pg:rev4_1}In the example in Section~\ref{subsec:ngn_application} we used the forward-filtering backward-sampling algorithm \citep[e.g.][]{Fru94} but other methods that rely on banded or sparse matrix algorithms will be more efficient in higher-dimensional applications; see, for example, \citet{CJ09}. The full conditional distributions for the transformed partial autocorrelation matrices $\matr{A}_1, \ldots, \matr{A}_m$ are non-standard. Given a truncation level of $H$, each $\matr{A}_i$ contains $H^2$ parameters and so it is not generally feasible to update the whole matrix in a single Metropolis-Hastings step. We therefore propose dividing the elements of each $\matr{A}_i$ into blocks of length $b$ and updating the blocks one-at-a-time. In the application in Section~\ref{subsec:ngn_application}, we found mixing to be greatly improved by using Metropolis-adjusted Langevin (MALA) steps rather than Gaussian random walks. 

Given a large enough fixed truncation level $H \le \lceil \varphi(p) \rceil - 1$, the MGP prior increasingly shrinks the factor loadings in columns with higher indices if the data suggest their likelihood contribution is negligible. At least conceptually, these columns can be discarded. If there are $m^{[i]}$ discardable columns on iteration $i$ of the MCMC sampler, the effective number of factors is defined as $k^{\ast\, [i]}=H - m^{[i]}$. The posterior for the effective number of factors can then be used as a proxy for the posterior for $k$. There are various truncation criteria for deciding which columns can be omitted. \citet{BD11} consider as inactive any factor whose loadings are all within $\epsilon$ of zero for some chosen threshold, say $\epsilon=10^{-4}$. \citet{SC20} suggest a more interpretable criterion which, unlike the choice of threshold $\epsilon$, is unaffected by the scale of the data and its dimension. Denote by $\matr{\Lambda}_{k^{\ast}}$ the factor loadings matrix obtained by discarding the columns of $\matr{\Lambda}$ from $k^{\ast}+1$ onwards. The basic idea is that if $\matr{\Omega}_{k^{\ast}} = \matr{\Lambda}_{k^{\ast}} \matr{\Lambda}_{k^{\ast}}^\T + \matr{\Sigma}$ can explain $100T$\% of the total variability of the data for some $T \in (0, 1)$, then one decides there are $k^{\ast}$ active factors. The choice of proportion $T$ then replaces the choice of threshold $\epsilon$. \label{pg:rev1_6}This is the truncation criterion we adopt in the applications in Section~\ref{sec:applications}.

When the truncation level $H$ is fixed, an alternative to Gibbs sampling is to implement off-the-shelf computational inference using standard probabilistic programming software. Code for the applications in Section~\ref{sec:applications}, written in Stan \citep[][]{CGH17}, is given in the Supplementary Materials. Stan uses Hamiltonian Monte Carlo to perform joint updates of all unknowns and so tends to converge faster and mix better than a Gibbs sampler. The caveat is the necessity to fix $H$; although this can be addressed in a bespoke implementation of Gibbs sampling (see Section~\ref{subsec:gibbs_unknown_k} below), Stan cannot be customised in this way. 

\subsection{\label{subsec:gibbs_unknown_k}Adaptive Gibbs sampler}

Especially in applications where $p$ is large, the number of effective factors is often considerably less than the Ledermann bound $\varphi(p)$. Therefore fixing a truncation level $H$ which is less than, but in the vicinity of, $\varphi(p)$ can be computationally inefficient. In the literature on infinite factor models, a common pragmatic solution is to use an adaptive Gibbs sampler which tunes the truncation level $H$ as it proceeds. Adaptation occurs at iteration $i$ with probability $p(i)=\exp(\alpha_0+\alpha_1 i)$ where $\alpha_0 \le 0$ and $\alpha_1 < 0$ such that the probability of adaptation decreases over the course of the simulation. This is necessary to satisfy the diminishing adaptation condition of \citet{RR07}. 

During an adaptation step, $k^{\ast\, [i]}$ is compared to the current value of $H$. The basic idea is to delete any inactive factors or to add an extra factor if all $H$ factors are active. In the static model, implementation is straightforward. If $k^{\ast\, [i]}<H$, $H$ is reduced to $k^{\ast\, [i]}$ and any inactive factors are deleted along with the corresponding columns of $\matr{\Lambda}$ and components of $\vec{\varrho}$ and $\matr{\Psi}$. If $k^{\ast\, [i]}=H$ and $H < \lceil \varphi(p) \rceil - 1$, $H$ is increased by 1 then an extra factor, column of factor loadings, and component of $\vec{\varrho}$ are sampled from their priors. In the dynamic model, the adaptation step is slightly more involved because of the additional complexity in the distribution of the latent factors. Full details of the adaptive Gibbs samplers for both static and dynamic models are given in the Supplementary Materials, along with R code for the applications in Section~\ref{sec:applications}. \label{pg:rev2_8}We also provide details on how the posterior output can be post-processed to obtain samples under a parameterisation in which the factor loadings matrix is identified by the PLT constraint. This is incredibly valuable for parameter interpretation and the assessment of MCMC diagnostics.



\section{\label{sec:applications}Applications}

\subsection{\label{subsec:simulation_expt}Simulation experiment}

\subsubsection{Simulation settings}
In order to investigate posterior sensitivity to the prior specification in the idealised setting in which we know the data were generated from a factor model with known shared variation matrix $\matr{\Delta}$, we carried out a series of simulation experiments. In the simplified setting in which $\matr{\Sigma}=\sigma^2 \matr{I}_p$, denote by $\beta$ the proportion of the total variation in $\matr{\Omega} = \matr{\Delta} + \matr{\Sigma}$ that is explained by the common factors, that is, $\beta=\trace(\matr{\Delta}) / \trace(\matr{\Omega})$. Then for given $\beta$ and $\matr{\Delta}$, the corresponding value of $\sigma^2$ can be computed as $\sigma^2=(1-\beta) \trace(\matr{\Delta}) / (k \beta)$. Three different combinations of $p$ and $k$, representing different degrees of dimension reduction, were considered: $(p=24, k=6)$, $(p=48, k=9)$ and $(p=72, k=9)$ so that $k/p$ was equal to 25\%, 18.75\% and 12.5\%, respectively. For each combination of $p$ and $k$, we also considered three values for $\beta$, $\beta \in \{0.9, 0.95, 0.99\}$, giving nine sets of values for $\{(p, k), \beta\}$ in total. We refer to these as \emph{simulation settings}. Note that we only set $\matr{\Sigma}=\sigma^2 \matr{I}_p$ for the purposes of simulating the data; for inference, we still allow $\matr{\Sigma} = \diag(\sigma_1^2, \ldots, \sigma_p^2)$.

Under each simulation setting, we simulated 24 data sets of size $n=50$ based on an exponential correlation matrix for $\matr{\Phi}$, taking $\matr{\Delta}$ to be the closest rank-$k$ matrix to $\matr{\Phi}$ (see Proposition~\ref{prop:min_distance}). In the simulation of each data set, in order to fix the value of $\matr{\Delta}$, we simulated $p$ pairs of coordinates $\vec{x}_j = (x_{j1}, x_{j2})^\T$, $j=1,\ldots,p$, by sampling uniformly at random from the unit square, and then set the expected shared variation matrix equal to $\matr{\Phi}$ with $(j,k)$ element $\phi_{jk}=\exp(-\|\vec{x}_j - \vec{x}_k \|_2 / \vartheta)$ where the length-scale parameter was equal to
\begin{equation*}
\vartheta =
\begin{cases}
-1 / \log(0.05), \quad &\text{for data sets $1,\ldots,6$;}\\
-1 / \log(0.10), \quad &\text{for data sets $7,\ldots,12$;}\\
-1 / \log(0.15), \quad &\text{for data sets $13,\ldots,18$;}\\
-1 / \log(0.20), \quad &\text{for data sets $19,\ldots,24$.}\\
\end{cases}
\end{equation*}
For a distance of 1 in the unit square, the correlations therefore ranged from 0.05 to 0.20.

\subsubsection{Prior distributions}
Each data set generated under each simulation setting was used to update five prior distributions for $\matr{\Lambda}$. Three are structured matrix normal priors, with different assumptions about $\matr{\Phi}$, and two are widely used priors in factor analysis which allow inference on the number of factors:
\begin{mydescription}
\item[SN (expcov.)] A structured matrix normal prior in which $\matr{\Phi}$ was taken to have the same form as that used to simulate the data, that is, an exponential correlation matrix. The length-scale parameter was taken to be unknown and assigned the distribution $\log(\vartheta) \sim \norm(0, 1)$. 
\item[SN (fixed)] A structured matrix normal prior in which $\matr{\Phi}$ was taken to have the same form as that used to simulate the data, that is, an exponential correlation matrix. The length-scale parameter was fixed at the value used to simulate the data.
\item[SN (exch.)] A structured matrix normal prior in which $\matr{\Phi}$ was taken to have a different form to that used to simulate the data, specifically we took $\matr{\Phi}=(1-\vartheta)\matr{I}_p + \vartheta \matr{J}_p$, $-1/(p-1) < \vartheta < 1$, representing a two-parameter exchangeable matrix with trace fixed at $p$. The correlation parameter was taken to be unknown and assigned the distribution $\logit(\check{\vartheta}) \sim \norm(0, 1)$ where $\check{\vartheta} = \{1 + (p-1) \vartheta \} / p$.
\item[MGP] The original (exchangeable) multiplicative gamma process prior described in \citep[][]{BD11}, which is a global-local prior. We took $a_1=1$ and $a_2=2$ in the multiplicative gamma process for the global precision parameters and $\nu=3$ in the distribution for the local precision parameters.
\item[CUSP] A cumulative shrinkage process prior \citep[][]{LDD20}, which is a slab-and-spike prior. We took $a_{\theta}=b_{\theta}=2$ in the distribution for the slab, $\theta_{\infty}=0.05$ for the position of the spike and $\alpha=5$ in the cumulative stick-breaking construction controlling the probability assigned to the spike.
\end{mydescription}
For the three structured matrix normal priors, we chose $a_1=1$ and $a_2=2$ in the multiplicative gamma process for the diagonal elements of $\matr{\Psi}$. In all cases, the idiosyncratic variances were assigned the prior $1 / \sigma_j^2 \sim \gamma(1, 0.3)$ independently for $j=1,\ldots,p$.

\subsubsection{\label{subsubsec:simulation_results}Analysis and results}
All analyses were run using an adaptive Gibbs sampler, allowing adaptation after 500 iterations and setting the parameters in the diminishing adaptation condition at $\alpha_0=-1$ and $\alpha_1=-5 \times 10^{-4}$. The chains were run for 10K iterations, following a burn-in of 10K, and thinned to retain every fifth sample to reduce computational overheads. For the three structured matrix normal priors and the original MGP prior, we applied the truncation criterion of \citet{SC20}, retaining as active enough factors to explain at least $T=0.95$ of the total variation in the data.

\begin{table}[tb]
\begin{center}
\caption{Summaries of the simulation experiment when $\beta=0.99$.}\label{tab:simulation_99}%
\begin{tabular}{@{}ccrrrr@{}}
\toprule
Prior &$(p, k)$ &\multicolumn{2}{c}{$\E\{ d_G(\matr{\Omega}, \matr{\Omega}_0) | \vec{y}\}$} &\multicolumn{2}{c}{$\E(k^\ast | \vec{y})$}\\ 
\cmidrule(lr){3-4} \cmidrule(lr){5-6}
 & &Median &IQR &Median &IQR\\
\midrule
   SN (expcov.) &(24, 6)       &15.88        &0.38      &6.34      &0.56\\ 
   SN (fixed) &(24, 6)       &15.78        &0.39      &6.30      &0.89\\ 
   SN (exch.) &(24, 6)       &15.93        &0.39      &6.63      &0.69\\ 
   MGP &(24, 6)       &16.02        &0.45      &6.27      &0.86\\ 
   CUSP &(24, 6)       &16.25        &0.36      &3.13      &0.57\\ 
   SN (expcov.) &(48, 9)       &21.25        &0.58      &7.54      &1.95\\ 
   SN (fixed) &(48, 9)       &21.18        &0.65      &7.32      &1.47\\ 
   SN (exch.) &(48, 9)       &22.20        &1.04      &5.96      &2.02\\ 
   MGP &(48, 9)       &22.15        &0.64      &6.41      &2.32\\ 
   CUSP &(48, 9)       &22.86        &0.84      &3.00      &0.23\\ 
   SN (expcov.) &(72, 9)       &23.89        &1.03      &7.76      &1.38\\ 
   SN (fixed) &(72, 9)       &23.69        &1.37      &7.85      &1.55\\ 
   SN (exch.) &(72, 9)       &25.81        &1.87      &6.15      &0.92\\ 
   MGP &(72, 9)       &24.99        &0.94      &7.21      &1.81\\ 
   CUSP &(72, 9)       &27.89        &1.07      &3.00      &0.00\\ 
\bottomrule
\end{tabular}
\end{center}
\end{table}

The (squared) geodesic distance between two symmetric positive definite matrices $\matr{A} \in \mathcal{S}_p^+$ and $\matr{B} \in \mathcal{S}_p^+$ is defined as
\begin{equation*}
d_G(\matr{A}, \matr{B})^2 = \left[ \sum_{i=1}^p \log\left\{ \lambda_i\left( \matr{A}^{-1} \matr{B} \right) \right\}^2 \right]
\end{equation*}
in which $\lambda_i(\matr{X})$ denotes the $i$th eigenvalue of $\matr{X}$ \citep[][]{LS19}. For simulation experiments in which $\beta=0.99$, Table~\ref{tab:simulation_99} reports the median and interquartile range across the 24 data sets of the posterior mean of $d_G(\matr{\Omega}, \matr{\Omega}_0)$ where $\matr{\Omega}_0$ denotes the value used to simulate the data. Corresponding Tables for the simulations where $\beta=0.95$ and $\beta=0.9$ are presented in the Supplementary Materials. We chose the distance metric $d_G(\cdot,\cdot)$ because, by construction, $\matr{\Omega}$ is close to singular and so the geodesic distance will be better suited than a metric, such as the Frobenius distance, which ignores the geometry of the space. In particular, it will give large values whenever $\matr{\Omega}$ and $\matr{\Omega}_{0}$ do not approximately span the same subspace of $\rnth{p}$, that is when their null spaces are not approximately the same. We also report the median and interquartile range for the posterior mean of the effective number of factors $k^\ast$.

Focusing first on the posterior mean of the geodesic distance $d_G(\matr{\Omega}, \matr{\Omega}_0)$, it is immediately clear from Table~\ref{tab:simulation_99} and the corresponding tables in the Supplementary Materials that both the medians and interquartile ranges across data sets are typically smallest when we use a prior in which $\matr{\Phi}$ is based on an exponential correlation matrix. As we would expect, this is especially true when the length-scale parameter in the prior is fixed at the value used in simulating the data. Nevertheless, the modest differences between posterior inferences under SN (expcov.) and SN (fixed) are reassuring; in the analysis of a real data set, though we may have strong prior beliefs about a suitable structure for $\E(\matr{\Delta})$, we are likely to be more uncertain about the values of the hyperparameters in that structure. We also observe that the gains from assuming the correct form for $\matr{\Phi}$ in this simulation experiment are greatest when the total variation in $\matr{\Delta}$ makes up a larger proportion of the total variation in $\matr{\Omega}$ (that is, when $\beta=0.99$ and $\beta=0.95$) and when the degree of dimension reduction is larger. This is as expected because in both cases, the prior for $\matr{\Lambda}$, and therefore $\matr{\Delta}$, is likely to impart more influence on the posterior for $\matr{\Omega}$. It is interesting to observe that when we adopt a structured matrix normal prior but assume an incorrect structure for $\matr{\Phi}$, the posterior mean for $d_G(\matr{\Omega}, \matr{\Omega}_0)$ still tends to be smaller than under the MGP or CUSP priors. This is likely to be because we simulated all data sets based on exponential correlation matrices for $\matr{\Phi}$ which have strictly positive diagonal elements. Under SN (exch.), we are shrinking $\matr{\Delta}$ towards a rank-reduced approximation to a two-parameter exchangeable matrix with a common (potentially positive) off-diagonal element whereas the prior expectation for $\matr{\Delta}$ under the MGP and CUSP priors is diagonal.

A discussion of the results concerning the posterior mean of the effective number of factors $k^\ast$ can be found in the Supplementary Materials.

\subsection{\label{subsec:finnish_birds}Co-occurrence of Finnish birds}

\subsubsection{Model and prior}
As discussed in Section~\ref{sec:introduction}, most of the literature on prior specification for factor models focuses on constructing a distribution for the factor loadings matrix that is exchangeable with respect to the order of the components in the observation vector. An exception is the \emph{structured increasing shrinkage (SIS) process} \citep[][]{SCD22}. This prior allows meta-covariates to inform the within-column sparsity structure of the factor loadings matrix in an infinite factorisation model. In this section, we re-examine the Finnish birds data set analysed in \citeauthor{SCD22} to illustrate the extra flexibility afforded by our approach. 

The Finnish bird data comprise information on the co-occurrence of the 50 most common bird species in Finland from 137 sampling areas in 2014 \citep[][]{LGHKL15}. The occurrences are arranged into a $n \times p$ binary matrix $\matr{Y}=(y_{ij})$ where $y_{ij}=1$ if bird species $j$ was observed in area $i$ and $y_{ij}=0$ otherwise. Information available to explain the variation in the mean $\vec{\mu}_i$ at each sampling area include a $n \times c$ matrix of environmental covariates $\matr{W}$ containing measurements on spring temperature and its square and a five-level habitat type (broadleaved forests, coniferous forests, open habitats, urban habitats, wetlands) so that $c=7$. Information available to structure the prior for the marginal variance $\matr{\Omega}$ include a phylogenetic tree, indicating the evolutionary relationships amongst the 50 species, and a $p \times q$ matrix of meta-covariates $\matr{X}$ consisting of two species traits: logarithm of typical body mass and a three-level migratory strategy (short-distance migrant, resident species, long-distance migrant). Following \citeauthor{SCD22}, we model species presence or absence using a multivariate probit regression model where $y_{ij} = \mathbb{I}(z_{ij}>0)$ and we assume the latent variables $\vec{z}_i = (z_{i1}, \ldots, z_{ip})^\T$ are such that
$\vec{z}_i = \matr{B}^\T \vec{w}_i + \matr{\Lambda} \vec{\eta}_i + \vec{\epsilon}_i$
for $i=1,\ldots,n$. Here $\vec{w}_i^\T$ is the $i^{\text{th}}$ row of $\matr{W}$, $\matr{B}=(\beta_{ij})$ is a $c \times p$ matrix of regression coefficients, and the idiosyncratic variances on the diagonal of $\Var(\vec{\epsilon}_i)=\matr{\Sigma}$ are set equal to 1 to prevent compensatory rescaling of $\matr{B}$ and $\matr{\Omega}=\matr{\Lambda}\matr{\Lambda}^\T + \matr{\Sigma}$. Our prior for the coefficients in $\matr{B}$ is identical to that of \citeauthor{SCD22} and described in the Supplementary Materials.

The SIS process prior for $\matr{\Lambda}$ is built on a series of assumptions of conditional independence. As a consequence, the expectation of the shared variation matrix $\matr{\Delta}$ is diagonal, irrespective of the priors assigned to its hyperparameters. Moreover, every factor loading $\lambda_{ij}$ has a spike-and-slab distribution in which the meta-covariates only enter through the logit of the spike probability. As such, there is no mechanism for incorporating the rich information on evolutionary relationships that is expressed through the phylogenetic tree.

In contrast, using the approach outlined in Section~\ref{subsec:prior_for_xi} we can structure the prior for the shared variation so that it incorporates the information from the phylogenetic tree in addition to the two meta-covariates. The former can be encoded as a matrix of phylogenetic distances between the 50 species, whose $(i,j)^{\text{th}}$ entry, say $d_{P,ij}$, is the sum of the branch lengths along the path between the leaves for species $i$ and $j$. This distance function is a metric over the set of leaves \citep[][Chapter 6]{Ste16}. Another measure of the dissimilarity between species can be obtained using the generalised distance function from the projection model in~\eqref{eq:distance}. As we only have binary response data, which is unlikely to be as informative as continuous data, we simplify~\eqref{eq:distance} by taking the matrix $\matr{\Theta}$ to be diagonal. This yields $d_{C,ij} = \left\{ \sum_{k=1}^4 (x_{ik} - x_{jk})^2 / \vartheta_k^2 \right\}^{1/2}$ as a metric for the distance between the meta-covariates for species $i$ and $j$, where $x_{i1}$, $x_{i2}$ and $x_{i3}$ are indicators for whether species $i$ is a short-distance migrant, resident or long-distance migrant, respectively, while $x_{i4}$ is the logarithm of typical body mass for species $i$. Since $d_{P,ij}$ and $d_{C,ij}$ are metrics over a set comprising the species and their associated traits, the sum is also a metric. We can therefore take as a generalised measure of distance between species $i$ and $j$, 
$d_{ij} = d_{C,ij} + d_{P,ij} / \vartheta_5$,
in which $\vartheta_i>0$ for $i=1,\ldots,5$. So that all the length-scale parameters $\vartheta_i$ operate on a comparable scale, the meta-covariates and branch lengths are standardised prior to calculating $d_{C,ij}$ and $d_{P,ij}$. Finally, we relate the generalised distance $d_{ij}$ to the among-row scale matrix $\matr{\Phi}$ by using the exponential correlation function, $\phi_{ij}=\exp(-d_{ij})$. We assume the length-scale parameters to be independent in the prior and assign distributions $\log \vartheta_i \sim \norm(0, 10)$.

Based on available phylogenetic and ecological information, this prior encourages shrinkage towards a rank-reduced approximation of an expected shared variation matrix $\E(\matr{\Delta})$ with higher correlations between types of bird that are more closely related in terms of their evolutionary history and species traits. Since we allow the length-scale parameters $\vartheta_i$ to be unknown we are also able to learn which variables are more important in explaining relationships between species. However, the structure for the expected shared variation matrix represented through $\matr{\Phi}$ is conjectural and so we elect to use the matrix-$t$ prior whose extra degree of freedom parameter $\varsigma$ allows the data to influence the degree of shrinkage. As discussed in Section~\ref{subsec:matrix_t_prior}, we induce a prior for $\varsigma > 4$ by specifying a distribution for $\check{\varsigma} = 1 / (\varsigma-4)  \in \mathbb{R}^+$. Specifically, we take $\check{\varsigma} \sim \expo(1)$ which implies that when the number of factors $k$ is 5, 10 and 15, the probability of the scale factor $s_k(\check{\varsigma})$ lying between 1 and $x$ is 0.75 for $x=7.8$, $x=12.9$ and $x=18.0$, respectively. For the diagonal elements in $\matr{\Psi}$, we assign a MGP prior~\eqref{eq:mgp} with $a_1=2$, $a_2=6$ and a maximum truncation point $H$ that satisfies $H \le \lceil \varphi(50) \rceil - 1 = 40$.


\subsubsection{Analysis and results}
We ran two chains, initialised at different starting points, using the adaptive Gibbs sampler discussed in Section~\ref{subsec:gibbs_unknown_k} and another two chains using Stan's implementation of the (non-adaptive) Hamiltonian Monte Carlo sampler with a fixed truncation level of $H=10$. In the Gibbs sampler, adaptation was allowed after 500 iterations and, following \citet{SCD22}, the parameters in the diminishing adaptation condition were set at $\alpha_0=-1$ and $\alpha_1=-5 \times 10^{-4}$. After a burn-in of 20K iterations, each chain was run for a further 20K iterations, thinning to store every fifth draw in order to reduce computational overheads. Using the truncation criterion of \citet{SC20}, we retain as active enough factors to explain at least $T=0.999$ of the total variation in the data. The posterior mode (and median) for the number of effective factors in each case was 5 with 95\% credible interval $(3, 6)$. Comparing the output of identified parameters across the chains from both samplers using the usual graphical diagnostic checks gave no evidence of any lack of convergence. 

To gauge performance against another non-exchangeable prior, the model was also fitted under the SIS process prior using an adaptive Gibbs sampler, with the hyperparameters in the prior and tuning parameters in the sampler set in line with the specification from Section~5 of \citet{SCD22}. In this case, more or less all the posterior mass for the number of effective factors was stacked at 4. 

\begin{figure}[!t]
\centering
\subfloat[][]{\label{fig:log_lengthscale}\includegraphics[width=0.4\textwidth]{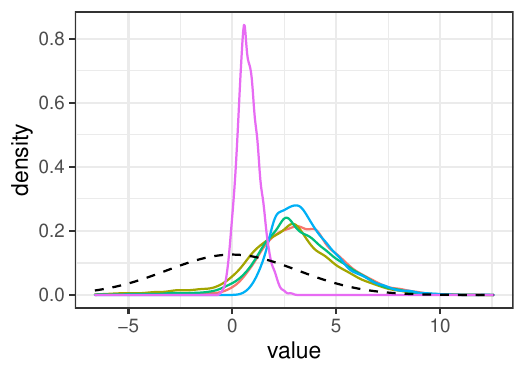}}\qquad
\subfloat[][]{\label{fig:dof_check}\includegraphics[width=0.4\textwidth]{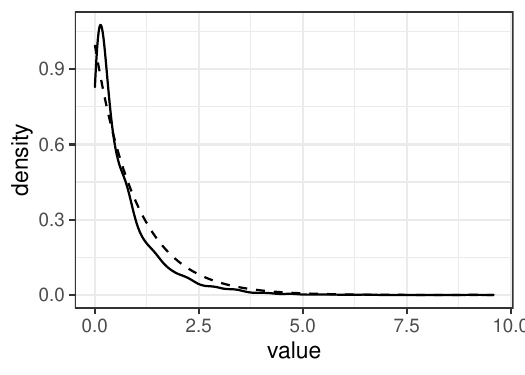}}%
\caption{Marginal posterior density for \protect\subref{fig:log_lengthscale} the logarithms of the length scale parameters $\vartheta_1$ (\textcolor{five1}{\protect\shortsolidline}), $\vartheta_2$ (\textcolor{five2}{\protect\shortsolidline}), $\vartheta_3$ (\textcolor{five3}{\protect\shortsolidline}), $\vartheta_4$ (\textcolor{five4}{\protect\shortsolidline}), $\vartheta_5$ (\textcolor{five5}{\protect\shortsolidline}) and \protect\subref{fig:dof_check} the transformed degree of freedom parameter $\check{\varsigma}$. Also shown are the priors (\protect\shortdashedline).}
\label{fig:posterior_hyperparameters}
\end{figure}

The marginal prior and posterior densities for the logarithms of the length-scale parameters $\vartheta_1,\ldots,\vartheta_5$ and the transformed degree of freedom parameter $\check{\varsigma} = 1 / (\varsigma-4)$ under the structured matrix-$t$ prior are displayed in Figure~\ref{fig:posterior_hyperparameters}. Absence of a relationship between a component of the generalised distance metric and the expected shared variation between species would be indicated by an infinite value for the corresponding length-scale parameter. Therefore the concentration of all posterior densities in Figure~\ref{fig:log_lengthscale} over a finite interval near the origin highlights the role played by the generalised distance metric in explaining the structure of the shared variation matrix. It is also clear that the phylogenetic distances, whose length-scale parameter is $\vartheta_5$, have the greatest influence. For the transformed degree of freedom parameter $\check{\varsigma}$ in Figure~\ref{fig:dof_check}, the similarity between the prior and posterior indicate that the information learned from the data is modest. Nevertheless, the shift in the position of the mode from 0 in the prior to around 0.1 in the posterior suggests that the data support less shrinkage towards the expected shared variation matrix than would be implied by a matrix normal prior for which $\check{\varsigma}=0$.

\begin{figure}[!t]
\centering
\includegraphics[width=0.95\textwidth]{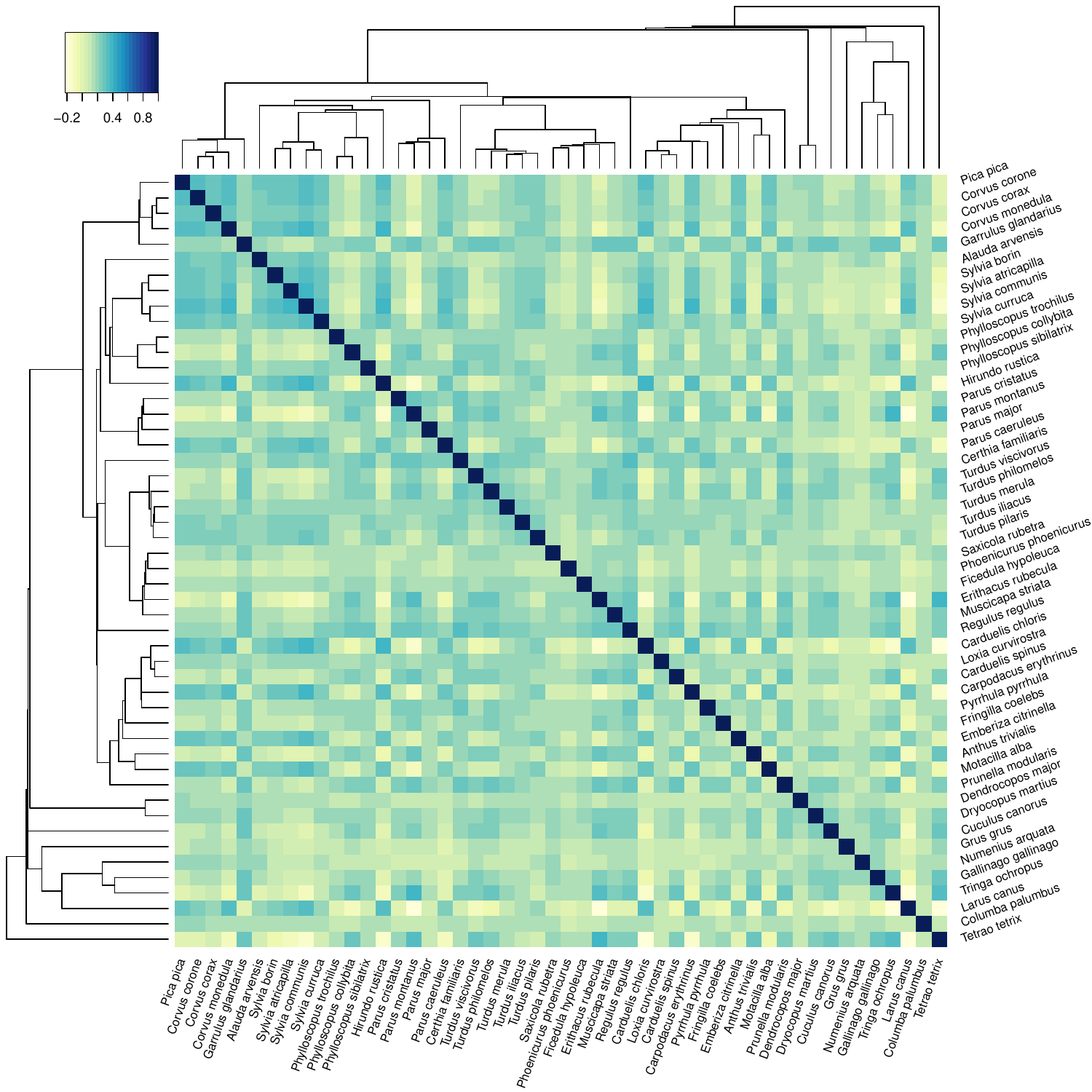}
\caption{Mean of the posterior for the marginal correlation matrix under the structured matrix-$t$ prior. The species are ordered according to their phylogenetic tree which is visualised in the margins.}
\label{fig:correlation_matrix}
\end{figure}

Further qualitative evidence of the importance of the phylogenetic information on the structure of the marginal covariance matrix $\matr{\Omega}=(\omega_{ij})$ is demonstrated in Figure~\ref{fig:correlation_matrix} which shows the mean of the posterior for the marginal correlation matrix $\matr{S}_{\Omega}^{-1} \matr{\Omega} \matr{S}_{\Omega}^{-1}$, where $\matr{S}_{\Omega} = \diag(\surd \omega_{11}, \ldots, \surd \omega_{pp})$.
The order of the 50 bird species is based on their phylogenetic tree, which is shown in the margins of the plot. There is clear structure in the matrix with several groups of species that share a recent common ancestor displaying strong positive correlations. This includes the species in the first ten rows, which correspond to birds from the Corvoidea superfamily and the Sylviidae family; the species in rows 19 to 30, which correspond largely to birds from the Musicapoidea superfamily; and the species in rows 42 to 47, which correspond mostly to birds from the Scolopacidae superfamily. It is also clear that most of the rows and columns predominated by near-zero correlations correspond to basal species, like \textit{Columba palumbus} and \textit{Grus grus}. Nevertheless, the structure of the matrix is clearly not dictated by evolutionary proximity alone. Interestingly, the corresponding plot from the analysis under the SIS process prior appears to be very sparse; see the Supplementary Materials. At least in part, this can be attributed to shrinkage of $\matr{\Delta}$ towards its prior mean, which is a diagonal matrix.

In order to provide a more formal comparison between the two model-prior combinations, we consider measures of goodness-of-fit and predictive performance. In the former case, we follow \citet{SCD22} by computing the pseudo marginal likelihood (PML), defined as the product of conditional predictive ordinates \citep[][]{GD94}. In terms of predictive performance, we consider the Brier score, which is a widely used proper scoring rule for categorical variables, calculated using a 4-fold cross-validation approach \citep[e.g.][]{GR07}. In our case, the score is positively oriented so that large values indicate better performance. The results are shown in Table~\ref{tab:model_comparison} from which we can conclude that the combination of the factor model and structured matrix-$t$ prior gives the best-fit to the data and, more markedly, the best predictive performance. Further details on the calculation of the PML and Brier score can be found in the Supplementary Materials.

\begin{table}[tb]
\begin{center}
\caption{Goodness-of-fit and predictive performance of the two model-prior combinations.}\label{tab:model_comparison}%
\begin{tabular}{@{}llll@{}}
\toprule
 & Structured matrix-$t$ & SIS process\\
\midrule
Log PML           &\textbf{-2831.6}  &-2936.5 \\
Brier score       &\textbf{-0.29120} &-0.85396\\
\bottomrule
\end{tabular}
\end{center}
\end{table}

\subsection{\label{subsec:ngn_application}Hourly demand for natural gas}

\subsubsection{Model and prior}

To illustrate the use of our methodology in the context of a dynamic model, we consider an application to modelling the hourly demand for natural gas. In the UK, gas in the national transmission system is transported to individual customers through eight regional distribution networks which are responsible for maintaining a reliable gas supply to customers at all times. This requires accurate short-term forecasts of the hourly demand for gas at individual points in the network called offtakes, where gas is taken off to supply locally. We consider data from a single offtake site in a regional distribution network in the North of England. Hourly gas demand data are available for the period from May 2012 to June 2015 along with the average daily temperature at the offtake site. In addition to the weather, the other main factors that are known to influence the mean level of residential demand for gas are seasonal and calendar effects \citep[]{Sol12}. We therefore additionally incorporate covariates that represent the day of the year, day of the week and whether or not a day is a public holiday.

The measurement of the demand for gas at an hourly resolution makes construction of a model challenging because the underlying dynamics of the process are likely to vary over a much longer time scale. For example, consider a dynamic linear model for hourly gas demand that allows for calendar and weather effects and incorporates a time-varying local level. Suppose that this local level is modelled as evolving according to a random walk or autoregression. Deviations from behaviour that would be considered typical given the calendar and weather effects would likely persist over days, rather than hours, and so hourly evolution of the local level would demand a near-zero innovation variance. Similarly, if calendar effects such as the day of the week, were regarded as dynamic, hourly evolution would demand vanishingly small innovation variances.

Motivated by these concerns, we define a \emph{gas-day-vector} as a longitudinal series of measurements of the log-transformed demand for gas over 24 hours, starting at 07:00, which is the beginning of a gas-day. Denote by $\vec{y}_t = (y_{t1}, \ldots,y_{t,24})^\T$ the gas-day-vector on day $t$ so that $y_{t,h}$ denotes the log-transformed demand for gas at hour $h$ of gas-day $t$. We elect to model the gas-day-vectors using a dynamic factor model with observation equation
$\vec{y}_t = \vec{\mu}_t + \matr{\Lambda} \vec{\eta}_t + \vec{\epsilon}_t$, where $\vec{\epsilon}_t \sim \norm_p(\vecn{0}, \matr{\Sigma})$,
$t=1,\ldots,n$. For simplicity in this illustrative application, we model evolution of the factors using a first order vector autoregression,
$\vec{\eta}_t = \matr{\Gamma} \vec{\eta}_{t-1} + \vec{\zeta}_t$, where $\vec{\zeta}_t \sim \norm_k(\vecn{0}, \matr{\Pi})$.
Due to the stability of the demand for gas over the time period in question, it is reasonable to assume that the departures from the time-varying mean $\vec{y}_t - \vec{\mu}_t$ follow a stationary process. We therefore constrain the stationary variance of the $\vec{\eta}_t$ to be $\matr{I}_k$ and take the initial distribution to be $\vec{\eta}_0 \sim \norm_k(\vecn{0}, \matr{I}_k)$. Reparameterising the model for the factors in terms of the transformed partial autocorrelation matrices yields a single unconstrained square matrix $\matr{A}=(\alpha_{ij})$ with $\matr{\Gamma}=(\matr{I}_k+\matr{A}\matr{A}^\T)^{-1/2} \matr{A}$ and $\matr{\Pi}=(\matr{I}_k+\matr{A} \matr{A}^\T)^{-1}$. As discussed in Section~\ref{subsec:dyn_param_expansion}, we assign a rotatable prior to $\matr{A}=(\alpha_{ij})$, and therefore $\matr{\Gamma}$, by taking $\alpha_{ij} \sim \norm(0, 1)$. 

The variances of the specific factors $\vec{\epsilon}_t$ are assigned the prior $1 / \sigma_{j}^2 \sim \gamma(3.1, 2.1)$ independently for $j=1,\ldots,p$ which ensures that the distributions have finite variance. The time-varying mean $\vec{\mu}_t$ is modelled as $\vec{\mu}_t = \matr{B}^\T \vec{w}_t$ in which $\vec{w}_t^\T = (w_{t1}, \ldots, w_{tc})$ is row $t$ of an $n \times c$ matrix of daily covariates $\matr{W}$. The columns of $\matr{W}$ allow for an intercept, a non-linear effect of deviations from seasonal average temperature and fixed effects for the day-of-the-week, the occurrence of a public holiday, and the day-of-the-year, which is represented as a truncated Fourier series with six harmonics. Full details of the model for $\vec{\mu}_t$ are given in the Supplementary Materials along with a description of the associated priors. 


In order to reflect both the persistence of above or below average demand for gas and the longitudinal but circular nature of a gas-day-vector, a sensible model for the inverse of the expected shared variation matrix $\matr{\Xi}=\matr{\Phi}^{-1}$ would be the precision matrix of a stationary circular autoregressive process of order one \citep[][]{HR10}. We therefore take $\matr{\Xi}=(\xi_{ij})$ to be a tridiagonal Toeplitz matrix with corners; that is, $\xi_{i,i+1}=\xi_{i+1,i}=-\vartheta/2$ for $i=1,\ldots,p-1$, $\xi_{1p}=\xi_{p1}=-\vartheta/2$ and $\xi_{ii}= 1$ for $i=1,\ldots,p$ where we assume $\vartheta \in [0, 1)$ and take $\logit(\vartheta) \sim \norm(0, 2)$. Exploratory analysis using data from another offtake site suggest that this structure is sensible based on the inverse of the sample covariance matrix. Given that the existing model is already rather complex, we chose to adopt a matrix normal prior for $\matr{\Lambda}$ in this example. The diagonal elements in the among-column scale matrix $\matr{\Psi}$ are given a MGP prior~\eqref{eq:mgp}, with $a_1=2$, $a_2=3$ and a maximum truncation point of $H = \lceil \varphi(24) \rceil - 1 = 17$.

\subsubsection{Analysis and results}
We ran two chains, initialised at different starting points, using the adaptive Gibbs sampler discussed in Section~\ref{subsec:gibbs_unknown_k} and another two chains using the (non-adaptive) Hamiltonian Monte Carlo sampler with a fixed truncation level of $H = \lceil \varphi(24) \rceil - 1=17$. After a burn-in of 50K iterations, each Gibbs chain was run for a further 200K iterations, thinning to retain every 100th draw in order to reduce computational overheads. Adaptation was allowed after 5K iterations and the parameters in the diminishing adaptation condition were set at $\alpha_0=-1$ and $\alpha_1=-5 \times 10^{-5}$. The Hamiltonian Monte Carlo sampler was run for 20K iterations after a burn-in of the same length, thinning the output to retain every 10th draw. As in the Finnish birds example, we retain as active enough factors to explain at least $T=0.999$ of the total variation in the data. The posterior mode (and median) for the number of effective factors in each case was 10 with 95\% credible interval $(9, 12)$. Comparing the output of identified parameters across the chains for the two samplers using the usual graphical diagnostic checks gave no evidence of any lack of convergence. 

\begin{figure}[!t]
\centering
\includegraphics[width=0.95\textwidth]{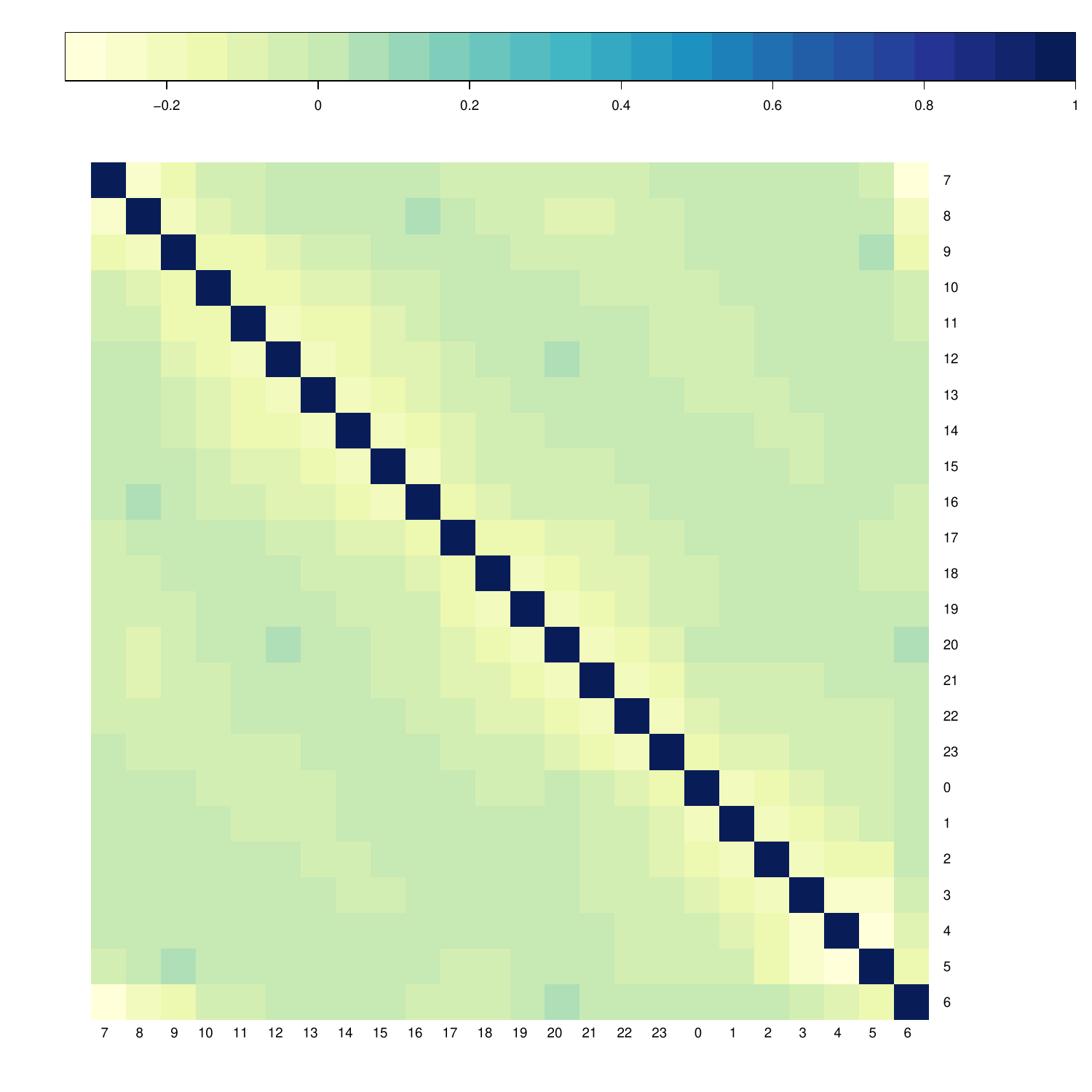}
\caption{Mean of the posterior for the marginal standardised precision matrix. The labels indicate hour of the day in a 24-hour clock.}
\label{fig:ngn_std_precision_matrix}
\end{figure}

Let $\matr{\Upsilon}=\matr{\Omega}^{-1}=(\upsilon_{ij})$ denote the marginal precision matrix of the process and let $\matr{S}_{\Upsilon} = \diag(\surd \upsilon_{11}, \ldots, \surd \upsilon_{pp})$. Figure~\ref{fig:ngn_std_precision_matrix} shows the posterior mean for the standardised precision matrix $\matr{S}_{\Upsilon}^{-1} \matr{\Upsilon} \matr{S}_{\Upsilon}^{-1}$. It is clear that its structure is reasonably consistent with the tridiagonal Toeplitz matrix with corners on which the prior for $\matr{\Delta}^{-1}$ is centered. However, there are some deviations, most notably another one or two bands of non-zero elements below the subdiagonal and above the supradiagonal. There is also some evidence of at least a partial band in the vicinity of $\upsilon_{i,i+12}$ for $i=1,\ldots,12$. This may be due to people switching their heating on twice per day, at around 7:00 in the morning and around 19:00 in the evening. This picture is reinforced by the posterior for the identified factor loadings matrix $\IdLambda$, shown in the Supplementary Materials, many of whose columns display a double-hump shape with smaller loadings towards the middle of the gas-day. As remarked in Section~\ref{subsec:enforcing_stationarity}, in order to simplify the geometry of the stationary region for stationary dynamic factor models, the autoregressive coefficient matrices in the state equation are often assumed to be diagonal. However, this example provides considerable evidence in favour of our fully flexible approach, with the posterior probabilities $\Pr(\tilde{\gamma}_{i,jk} >0 | \vec{y}_{1:n})$ for many off-diagonal elements being close to zero or one. 



\begin{figure}[!t]
\centering
\includegraphics[width=0.85\textwidth]{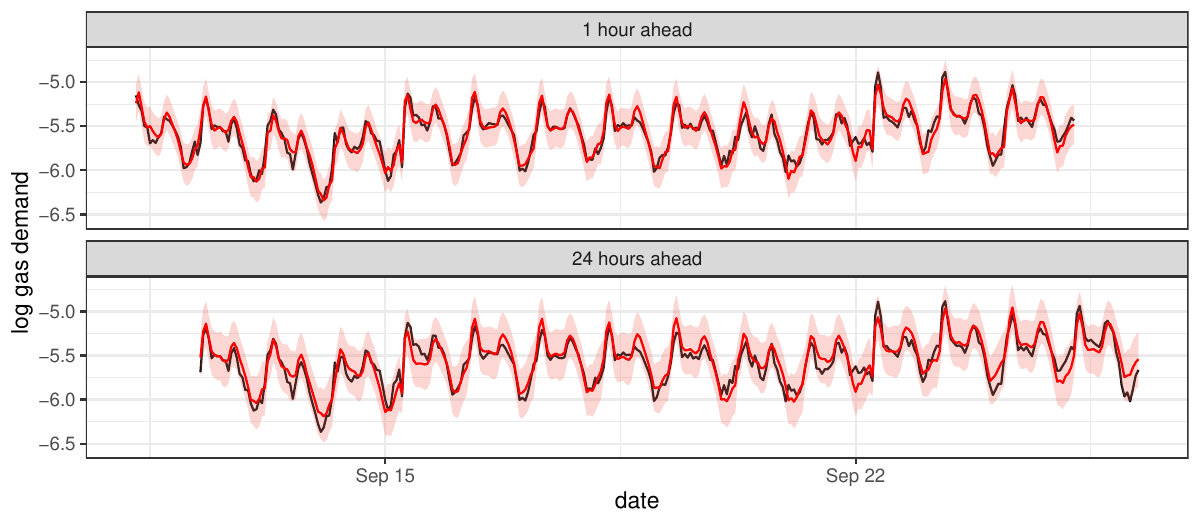}
\caption{For the first 5\% of times in the hold-out period, means (\textcolor{red}{\protect\shortsolidline}) and 95\% equi-tailed credible intervals (\colorbox{red!30!white}{\textcolor{red!30!white}{\protect\shortsolidline}}) for the one-step ahead and 24-step ahead posterior predictive distributions. The observed data are also shown (\protect\shortsolidline).}
\label{fig:ngn_forecast_first_nolegend}
\end{figure}

In order to assess the forecasting performance of the model, the number of factors was fixed at $k=10$ and the model was refitted, holding back the last 25\% of observations as test data. In keeping with the short-term horizon of interest, we considered forecasting $h=24$ hours ahead and, for comparative purposes, $h=1$ hour ahead. Application of the standard forward filtering and forecasting recursions for (multivariate) dynamic linear models are not appropriate here as they do not allow within-day updates or forecasts. We therefore modify the standard forward filter so that the time-step is hours, rather than days, and at hours $2,\ldots,24$ within each gas-day, we perform a partial update, comprising an observation step but no prediction step. The forecasting algorithm is similarly modified to allow forecasts to be issued hourly. The full algorithms are given in the Supplementary Materials. By using the draws from the posterior obtained from the fit to the training data and sampling one-for-one from the $h$-step ahead predictive distributions, we obtain samples from the $h$-step ahead posterior predictive distributions. For $h=1$ and $h=24$, these are visualised in Figure~\ref{fig:ngn_forecast_first_nolegend} for the first 5\% of times in the hold-out period, along with the test data. An analogous figure in the Supplementary Materials shows the last 5\% of times. From these plots it is clear that the posterior predictive distributions are both accurate and precise over the forecast horizons of interest.

\section{\label{sec:discussion}Discussion}

We have proposed a class of structured priors for the loadings matrix of a Bayesian factor model with accompanying inferential algorithms. The novelty lies in the insight that the shared variation matrix $\matr{\Delta}=\matr{\Lambda}\matr{\Lambda}^\T$ is much more amenable to the specification of prior beliefs than the factor loadings matrix $\matr{\Lambda}$. The prior is based on a matrix normal or matrix-$t$ distribution for $\matr{\Lambda}$. Two important features are the choice of parametric structure for the among-row scale matrix $\matr{\Phi}$ and the increasing shrinkage prior for the diagonal among-column scale matrix $\matr{\Psi}$. The matrix $\matr{\Phi}$ characterises the conditional prior expectation of $\matr{\Delta}$. Parametric forms are widely adopted as models for covariance matrices in longitudinal studies, spatial statistics, Gaussian processes, and a host of other areas. By adopting a parametric form, informed by domain expertise, as a model for the expectation of $\matr{\Delta}$, we benefit from shrinkage towards a rank-reduced matrix that is close to that structure. In general, the number of factors in a factor model is not known \textit{a priori}. This is addressed through the increasing shrinkage process prior for $\psi_1, \ldots, \psi_k$ which allows columns of loadings whose likelihood contribution is negligible to be shrunk towards zero and discarded. A prior that encourages no more factors than are needed is helpful, particularly for prediction, because of the reduction in epistemic uncertainty that is afforded. At the cost of slightly more involved computational inference, the matrix-$t$ version of the structured prior also offers a degree of freedom parameter $\varsigma$ which allows the data to influence the degree of shrinkage of $\matr{\Delta}$ towards a rank-reduced matrix that is close to its mean.

\section*{Acknowledgements}

EH was supported by the EPSRC grant EP/N510129/1 via the Alan Turing Institute project ``Streaming data modelling for real-time monitoring and forecasting''. We are grateful to Michael Betancourt and Malcolm Farrow for conversations which have improved the manuscript. We are also grateful to two anonymous referees for their helpful comments and suggestions. The Finnish bird data are available publicly from \url{https://www.helsinki.fi/en/researchgroups/statistical-ecology/software/hmsc}.

\bibliography{journal_names_abbr,refs}

\end{document}